\documentclass{article}

\usepackage{graphicx}  
\usepackage{amsmath}   
\usepackage[compress]{cite}
\usepackage{amssymb}   
\usepackage{bm} 

\usepackage{mathrsfs}

\def\eq#1{{Eq.~(\ref{#1})}}

 \def\g{\sqrt{-g}\,}
\def\gu#1#2{{g^{#1#2}}}

\def\half{{\frac{1}{2}}}

\newcommand{\df}{\delta}

\def\sp#1#2{{\left[\pounds_{#1} {#2}\right]_{\rm std}}}

\def\cc{{cosmological\ constant}}

\newcommand{\LL}{Lanczos-Lovelock }

  \title{General Relativity from a Thermodynamic Perspective}
  \author{T. Padmanabhan\\
  IUCAA, Pune University Campus,\\
  Ganeshkhind, Pune- 411 007.\\
  {\small {email: paddy@iucaa.ernet.in}}
  }

  \date{ }  

\begin{document}
  
  \maketitle
\begin{abstract}
 The deep inter-relationship between gravitational dynamics and horizon thermodynamics suggests that gravity is an emergent phenomenon with its field equations having the same status as, say, the equations of fluid dynamics. I describe several additional results which strengthen this idea by establishing connections between  the gravitational dynamics in a bulk region of space and a  thermodynamic description in the boundary of that region.  In Sec.~\ref{sec:2}, I provide an ab-initio description of gravity
 in terms of $f^{ab} \equiv  \sqrt{-g}g^{ab}$ and its associated canonical momentum $N^c_{ab}$, and
 motivate the use of these variables.  I also review the  connection (already established in a recent work arXiv:1303.1535) between these variables ($f^{ab},N^c_{ab}$)   and the thermodynamic variables ($S,T$) associated with a null boundary. In Secs. 3--5, I motivate and derive the conserved currents associated with the vector fields from simple identities in differential geometry, without the use of any symmetry considerations. I then show that the conserved charge contained in a bulk region, associated with a \textit{specific} time evolution vector field, has  a direct thermodynamic interpretation as the gravitational heat density of the boundary surface. This, in turn, leads to the result that all static spacetimes maintain holographic equipartition; viz., in these spacetimes, the (naturally defined) number of  degrees of freedom in the boundary is equal to the number of  degrees of freedom in the bulk. More importantly, in a general, dynamic spacetime one can relate the rate of change of gravitational momentum to the difference between the number of bulk and boundary degrees of freedom. It is the departure from  holographic equipartition which drives the time evolution of the spacetime. The description also allows us to define a natural four momentum current associated with gravity and the corresponding energy contained in a bulk region. When the equations of motion hold, the total energy of the gravity plus matter system in a bulk region is equal to the boundary heat content. In Sec.~\ref{sec:6}, I extend these ideas  to null surfaces.  After motivating the need for an alternate description of gravity (if we have to solve the cosmological constant problem), I describe a thermodynamic variational principle based on null surfaces to achieve this goal. Once again, the concept of gravitational heat density arises naturally on the null boundaries, and the variational principle, in fact, extremises the total heat content of the matter plus gravity system. In the final Sec.~\ref{sec:7}, I describe several possible variations on this theme like the use of other equivalent Lagrangians (which differ from the Einstein-Hilbert Lagrangian by  total divergences) and the use of other time evolution vectors to define the conserved charges. Several implications of these results for the emergent paradigm of gravity are also described. 
\end{abstract}

\tableofcontents
  \section{Introduction and Summary of the results}\label{sec:1}
  
  The uncanny similarity between the laws of black hole dynamics  and classical thermodynamics, discovered in the seventies (e.g., \cite{a1,a2,a3,a4,a5}; for a review, see \cite{a6}), suggested a possible connection between gravitational dynamics and horizon thermodynamics. This  was further strengthened by the realization that black holes are in no way special and that thermodynamic parameters like temperature will be attributed to \textit{any} null surface by a class of observers who perceive the null surface as a horizon due to their state of motion, even in flat spacetime \cite{du}. This allows one to introduce the concept of local Rindler observers around any event in a spacetime \cite{a12,a17} just as one introduces freely falling observers around any event. The fact, that observers in different states of motion will attribute different thermodynamic features to null surfaces, introduces a new level of observer dependence into the physical theory (see Sec. 4 of \cite{a9}, Sec.  4.4 of \cite{a8}).
  
  Several further investigations regarding the dynamics of gravity have shown that this connection is far deeper than was originally suspected. In particular, we now know that the relation between horizon thermodynamics and gravitational dynamics transcends Einstein's theory of gravity and holds for a much wider class of models like, e.g., \LL\ models  \cite{llreview}. Such investigations have led to the emergent gravity paradigm in which  the  gravitational field equations have the same status as the equations of elasticity or fluid mechanics (for reviews, see e.g., \cite{a8,lessons}). The main purpose of this paper is to provide significant additional evidence for this paradigm within the context of general relativity. (I plan to address the generalization to \LL\ models of gravity in a future work.) In addition to new results, I will also paraphrase some known features of gravity in the language of the emergent paradigm because it provides better insights into these results. In fact, one of the significant fall-outs of the emergent gravity program was that it highlighted several peculiar features of general relativity which had previously not attracted the attention they deserved \cite{lessons}. Some of the results in this paper continue this tradition.

Since this is a somewhat long paper touching upon diverse aspects of gravitational dynamics, I will first  provide a summary of the key results of the paper. 
  
   I begin in the next section by essentially ``running a commercial'' for describing gravity using $f^{ab}\equiv \sqrt{-g}g^{ab}$ and the corresponding canonical momenta $N^c_{ab}$. These variables are, of course, not new (and, in fact, I will christen them as ``old variables'' \cite{a36,a37,a38}) and have appeared in literature from time to time (like e.g., in \cite{a45,a44}). Recently, we provided a fairly comprehensive discussion of these variables \cite{KBP} and demonstrated how the description of gravity takes a simple and elegant form in terms of these variables.  In Sec.~\ref{sec:2}, I will provide an ab-initio route to gravitational dynamics based on these variables. In addition to setting the stage, this discussion serves the following important purpose: 
  It introduces the notion of the  momentum density of the gravitational field and relates its variation to the Ricci tensor and --- through it --- to the dynamics. The key equation in this context is 
  \begin{equation}
   \pounds_q R_{ab} = -\nabla_c(\pounds_q N^c_{ab}) = \half \pounds_q \mathscr{F}_{ab}; \qquad \mathscr{F}_{ab} \equiv T_{ab} - \frac{1}{2} g_{ab}T
  \end{equation} 
(in units with $16\pi G =1=c$).
  This exact equation relates the variation of the gravitational momentum density $\pounds_q N^c_{ab} $ (in the form of a Lie derivative  along an arbitrary vector field $q^a$) to the corresponding variation in the Ricci tensor and --- through gravitational dynamics --- with the variation of the energy momentum tensor. As we shall see, many other results in the paper arise from the fact that the above equation allows us to relate  $f^{ab} \delta R_{ab} $ to $f^{ab} \delta N^l_{ab}$. The former quantity occurs in the variation of the surface term of the gravitational action and thus plays a crucial role  in: (i) the symplectic structure of gravity, (ii) the form of the Noether currents and (iii) the physics of the boundary surfaces, while the latter expression will occur ubiquitously throughout this paper. 
  
  I also review briefly the connection  between the variations of the dynamical variables ($f^{ab} \delta N^c_{ab}$, $ N^c_{ab} \delta f^{ab}$) and the variations of the thermodynamic variables ($S\delta T, T\delta S$). Previous work \cite{KBP} has established that there is a one-to-one correspondence between these two, when evaluated on the null surfaces for a specific class of metric variations which preserve the null surface geometry. In the rest of the paper, we repeatedly encounter these variations --- in particular, $f^{ab} \delta N^c_{ab}$ --- and it is important to keep in mind the thermodynamic correspondence between this variation and $S \delta T$. 
  
  The fact that two such variations --- $S\delta T$ and $T\delta S$ --- exist and can be related to each other is not often emphasized and it is worthwhile to discuss this briefly in a familiar context. Consider the standard black hole thermodynamics in which the addition of a mass $\delta M$ changes the energy of the black hole by the standard relation $\delta M=\delta E = T \delta S$. While this relation is often taken for granted as ``natural'' without a second thought, it does contain a very peculiar feature: The absence of a $S\delta T$ term makes it   appear as though $T$ is kept constant while the process takes place which, of course, is not true since the change in $M$ changes both $T$ and $S$ of the black hole. In fact, the Schwarzschild black hole (with horizon area $A$) satisfies the relation (with $L_P^2 = (G\hbar / c^3) $ being the Planck area):
  \begin{equation}
   E = 2TS = \frac{1}{2} T \left(\frac{A}{L_P^2}\right)
   \label{ets}
  \end{equation} 
  with a crucial factor of 2 [and $(1/2)$] in the two equalities above. The second relation tells us that, if we attribute $N_{\rm sur} = A/L_P^2$ degrees of freedom to the horizon area $A$, then each degree of freedom carries exactly $(1/2) k_BT$ amount of energy. This, in turn, tells us that if we attribute $N_{\rm bulk} \equiv [E/(1/2)T]$ degrees of freedom with the bulk gravitational energy inside the horizon (``equipartition''), then the black hole is in holographic equipartition with $N_{\rm sur} = N_{\rm bulk}$. From \eq{ets}, it follows that 
  \begin{equation}
   \delta E = 2S\delta T + 2T \delta S  
  \end{equation} 
  with both  the variations contributing to the change in energy. However, there is an additional relation $S \propto M^2 \propto T^{-2}$ which is maintained during the variation that comes to our rescue, allowing us to express $\delta E$ either in terms of $\delta S$ or in terms of $\delta T$ alone:
  \begin{equation}
   \delta E = T\delta S = - 2S \delta T = - \frac{1}{2} \frac{A}{L_P^2} \delta T = - \frac{1}{2} N_{\rm sur} \delta T
\label{e4}
  \end{equation} 
  While the first equality is so widely discussed, the (equally valid) remaining relations are seldom emphasized in the literature! Equation (\ref{e4})
  tells us that addition of energy can also be thought of as resulting in an increase in the temperature while the number of surface degrees of freedom is held fixed, with the crucial minus sign indicating the negative specific heat of the gravitating system. While, in reality, both $S$ and $T$ change in the physical processes involving the horizon, they could be described in a complementary fashion, as either $T \delta S$ or as $ - (1/2) (A/L_P^2) \delta T$. 
  (The latter is, in fact, similar to the description of, say, heating a mono-atomic ideal gas with $\delta E = (3N/2) \delta T$; we do not change  $N$ while heating a fixed amount of gas.)
  In the general context of null surfaces and other boundaries, we will find that it is the latter interpretation involving $\delta T$ (which corresponds to $\delta N^a_{bc}$) that provides a more natural description. As a result, we will continuously encounter this variation in different guises while studying the boundary thermodynamics. Roughly speaking, $f^{ab}$ acts like an extensive variable in thermodynamics, while $N^c_{ab}$ acts like an intensive variable.

A closely related quantity is the total variation $\delta (f^{ab} N^c_{ab})$ which corresponds to $\delta (TS)$. This expression arises when we study the contribution of the boundary term in the Einstein-Hilbert action on a local Rindler horizon, and it occurs without the factor 2 noted earlier. I have discussed the physical significance of this term (and the difference between $TS$ and $2TS$) in my previous papers \cite{a40,a41}. The $TS$  essentially corresponds to the enthalpy (or heat content)\footnote{In usual thermodynamics, enthalpy $H$ is defined as $H=E+PV$ and satisfies the identity $E+PV-TS=\mu N$ where $\mu$ is the chemical potential. So for systems with $\mu=0$, we have the result $H=TS$, which is what we can call the heat content of the system. In this paper, I will use the term enthalpy to refer to just $TS$.}\label{pagerefone}  of the gravitational field which measures the difference between the energy and free energy of a finite temperature system, while $2TS$ measures the equipartition energy obtained by attributing $(1/2)k_BT$ to each of the $(A/L_P^2)$ degrees of freedom in an area $A$. The basic result in the context of black holes, written in two forms:
\begin{equation}
 2(TS)=2(M/2)=\frac{1}{2}T (A/L_P^2); \quad 
2(\mathrm{heat\ content}) =(\mathrm{equipartition\ energy})
\end{equation} 
will keep appearing in our discussions.

  In Sections 3--5, I will discuss the relationship between the above concepts, Noether currents and the gravitational dynamics. To do this consistently, I will first show (Sec.~\ref{sec:3}) how Noether currents can be thought of as arising purely from some, actually rather trivial, mathematical identities in differential geometry. Introducing them in such a manner allows certain flexibility and brings sharply into focus their connection with dynamics which I introduce later on. (On the other hand, if I obtain Noether currents in the more conventional manner, from the diffeomorphism invariance of the action, then it is hardly surprising that they possess dynamical content since the action functional already knows about the dynamics of the system.) While one can associate a conserved current with \textit{any} vector field in spacetime, the  vector fields related to the time evolution  are special. If we foliate the spacetime in the usual manner with $u_a$ denoting the unit normal to the $t=$ constant surfaces, then the two vector fields which are closely related to time evolution are the following: 
  \begin{equation}
   \xi^a \equiv N u^a; \qquad \zeta^a \equiv N u^a + N^a
  \end{equation} 
  where $N$ and $N^a \equiv h^a_b \zeta^b $ are the lapse and shift functions and $h^a_b=\delta^a_b+u^au_b$ is the projection tensor. The first one ($\xi^a$)  is related to the flow of proper time normal to the $t=$ constant surfaces and is parallel to the velocity vector $u^a$ of the fundamental observers. The second one ($\zeta^a$) also satisfies the condition $\zeta^a \nabla_a t = 1$ \textit{and} takes into account the shift of spatial coordinates between two spacelike hypersurfaces. In the past literature, $\zeta^a$ has been used and investigated in many different contexts quite widely. I will, however, show that one can obtain quite elegant and physically pleasing results from the Noether current and charge corresponding to the vector $\xi^a$. To begin with, I will show (Sec.~\ref{sec:4.1}) that the total Noether charge (associated with $\xi^a$) in any  bulk region $\mathcal{V}$ bounded by a constant lapse surface $\partial\mathcal{V}$,  is equal to the heat content of the boundary surface. Further,  twice the Noether charge gives the equipartition energy of the surface: 
  \begin{equation}
\int_\mathcal{V}\sqrt{h}\, d^3x\ u_aJ^a[\xi]
= \epsilon \int_{\partial\mathcal{V}}d^2x\ Ts; \quad
2\int_\mathcal{V}\sqrt{h}\, d^3x\ u_aJ^a[\xi]
= \epsilon \int_{\partial\mathcal{V}}\frac{\sqrt{\sigma}\, d^2x}{L_P^2}\frac{1}{2} (k_BT);
\end{equation}
where $\epsilon = \pm 1$.
The temperature and entropy density of a patch of area are defined using local Rindler observers (described later around \eq{flux3}).
This result holds in arbitrary, time evolving spacetimes and
provides a simple and direct interpretation of the Noether charge associated with one of the time development vectors. (For the sake of completeness, I also discuss briefly in Sec.~\ref{sec:7.3} the corresponding results for $\zeta^a$. Since there is a relative boost  between the observers moving along $\zeta^a$ and the fundamental observers moving along $\xi^a$, there are additional contributions related to the boost energy which mar the simple thermodynamic interpretation in the case of $\zeta^a$.) 
  
The above relation also allows one to study the concept of holographic equipartition 
\cite{cqg04,a20,a21}
in static spacetimes and --- more importantly --- relate the time evolution of the spacetime geometry to the departure from holographic equipartition. I first re-derive (Sec.~\ref{sec:4.2}) the holographic equipartition law in the form $N_{\rm sur} = N_{\rm bulk}$ for any static spacetime with natural definitions for the surface and bulk degrees of freedom. Further, I show that the time evolution of the spacetime geometry can be described in a rather elegant manner by the equation:
  \begin{equation}
\int_\mathcal{V} \frac{d^3x}{8\pi}\sqrt{h} u_a g^{ij} \pounds_\xi N^a_{ij} = -\int_\mathcal{V}\frac{d^3x}{8\pi}h_{ab}\pounds_\xi p^{ab}  =  \epsilon \frac{1}{2} k_B T_{\rm avg} ( N_{\rm sur} - N_{\rm bulk})
\end{equation} 
where $h_{ab}$ is the induced metric on the $t=$ constant surfaces, $p^{ab}$ is its conjugate momentum and $T_{\rm avg}$ is the average Davies-Unruh temperature  of the boundary of $\mathcal{V}$. On the left hand side, we again see the rate of change of gravitational momentum which, in thermodynamic language, is related to the change in the temperature. The above equation shows that this change is driven by the departure from holographic equipartition, and that time evolution will cease when $N_{\rm sur} = N_{\rm bulk}$. I stress that all these concepts depend on the choice of observers and their accelerations, etc. arising through the spacetime foliation. While the equations are generally covariant, they encode  observer dependence through the choice of the foliation and non-geometrical fields like the velocity of the observers. \textit{I emphasize that this is a feature and not a bug.} As I mentioned earlier, attribution of thermodynamic variables to a null surface \textit{is} observer dependent. Therefore, if we relate the dynamical evolution to the thermodynamical concepts,  different observers using different coordinate systems \textit{must} perceive, for example, the time evolution of the metric, differently. We know that the spacetime metric describing  a black hole is time independent in the Schwarzschild coordinates but not in, say, the synchronous coordinates; what we see here is fundamentally no different from this fact. 

A closely related result pertains to the definition and evaluation of the energy associated with gravity in a bulk region of space.  Defining energy for  gravity is a notorious activity that  evokes strong reactions from the cognoscenti. I will, nevertheless,  show that the description of gravity in terms of ($f^{ab}, N^c_{ab}$) provides a  natural way of associating a gravitational four-momentum current with every vector field. This current is, in fact, what is usually called the Noether current for the vector field in the literature except that we will now: (i) choose it for the specific time development vector $\xi^a$ and (ii) motivate it somewhat more physically. More importantly, we will show  that the total amount of gravitational energy contained in a bulk region is then exactly equal to the surface heat content when the equations of motion hold.  That is, we will show that for any self-gravitating system, which could be time-dependent and dynamically evolving, the total energy contained in a region $\mathcal{R}$ bounded by a constant lapse surface $\partial\mathcal{R}$, is given by
  \begin{equation}
 \int_\mathcal{R} d^3x\sqrt{h}u_a[P^a(\xi)+NT^a_bu^b]=\int_{\partial\mathcal{R}} d^2x \ Ts
\end{equation} 
  On the left hand side, the first term  is the suitably defined gravitational contribution to the energy density and the second term is the contribution from matter. While it is well known in the literature that the quasi-local energies (for a review, see e.g., \cite{qle}) for gravity are usually given by boundary integrals, the precise thermodynamic meaning we obtain is important in revealing the holographic nature of gravity.
  
  I stress that the validity of the  above result is \textit{not} restricted to static spacetimes, which --- in turn --- implies that this energy and the boundary heat content will evolve in time due to physical processes. By adopting the usual techniques used to obtain the symplectic form of gravity from Noether currents, one can easily determine the variation of this energy with respect to the chosen time development vector. I will show that this is given by 
  \begin{equation}
   \pounds_\xi H_{\rm grav} =  \int_{\partial\mathcal{R}} d^2 x \, \sqrt{\sigma}\, N r_a \left( T^{ab} \xi_b + \gu lm \pounds_\xi N^a_{lm}\right)
\end{equation} 
where $r_a$ is the normal to $\partial \mathcal{R}$.
 Of the two terms in the integrand, the first one is  due to the matter energy flux across the boundary. The second term, representing  the  gravitational sector, again has the variation of $N^a_{lm}$, which reinforces the idea that this term can be thought of as the change in the gravitational heat due to the processes at the boundary.  For example, in a matter-free spacetime (describing e.g., gravitational waves), we will have
 \begin{equation}
   \pounds_\xi H_{\rm grav} =  \int_{\partial\mathcal{R}} d^2 x \, \sqrt{\sigma}\,
 N r_a   \gu lm \pounds_\xi N^a_{lm}
\end{equation}
showing that the flow of energy of  gravitational waves in a bulk region is described by the surface term.
The physical meaning is again thermodynamic and is related to the change in the heat content of the boundary. 
  
  Up to this point in the paper, I have been faithful to the party line and have only attempted to connect the conventional description of bulk gravitational dynamics with boundary thermodynamics. I, however, believe that the \textit{so called \cc\ problem is a clear indication that we are ignoring the most significant clue we have about the nature of gravitational dynamics}: Gravity does not couple to changes in the  bulk energy density arising from the addition of a constant to the matter Lagrangian. I begin Sec.~\ref{sec:6.1} by showing that one cannot solve the \cc\ problem in any generally covariant theory in which the metric is varied as a dynamical variable, in an unrestricted manner in a local action principle. A possible alternative is to obtain the field equations of gravity as a consistency condition for the spacetime from  a variational principle involving some other dynamical variable. This is indeed possible (not only for Einstein's gravity but even for all \LL\ theories \cite{aseemtp}) by defining a thermodynamic functional on every null surface and demanding that 
this functional should be extremised simultaneously on all null surfaces. I connect up this formalism with the heat density of the null surfaces and relate it again to a Noether charge. In this case, the Noether current is evaluated for the null vector defining the congruence, which is analogous to  the time development vector in the case of spacelike surfaces discussed earlier. It turns out that the field equations for gravity in the absence of matter can be obtained by extremising the following functional on all null surfaces simultaneously:
\begin{equation}
 Q\equiv \frac{1}{16\pi}\int_{\lambda_1}^{\lambda_2} d\lambda \ d^2x\, \sqrt{\sigma}\, \left[g^{ij} \ell_a\pounds_\ell N^a_{ij} \right]
\end{equation}
Here, $\ell^a$ is the affinely parametrized null vector defining the congruence.
The occurrence of $\gu ij\pounds_\ell N^a_{ij}$, which --- as we said before, is related to $S\delta T$ --- suggests that this extremum principle can be given a thermodynamic interpretation.
This strengthens the physical interpretation of the variational principle and relates it to the corresponding concepts on spacelike and timelike boundaries.
  
  Right at the end (in Sec.~\ref{sec:7}) I discuss several other variations on the basic theme presented here. Among them, the possibility of using different Lagrangians, related to the Einstein-Hilbert Lagrangian by total divergences, and the possibility of using other time development vector fields (in particular $\zeta^a$) are of some interest. I show how the Noether potential gets modified by the addition of total divergences to the Einstein-Hilbert Lagrangian. In particular, it turns out that the results related to a spacelike surface, obtained with the time development vector $\xi^a$, do \textit{not} change when we use the Lagrangian with the addition of the $2K$ term at the boundary. (This does not happen for $\zeta^a$, which is possibly yet another reason to use $\xi^a$ in the formalism.) Corrections do appear when one uses the $\Gamma^2$ Lagrangian which I will describe briefly. Finally, I describe the results pertaining to the use of $\zeta^a$ rather than $\xi^a$. As mentioned earlier, this introduces an additional boost relative to the fundamental observers which complicates the interpretation. To a limited extent, this can be understood in terms of a modified acceleration $\zeta^a \nabla_a u^b$ instead of $N u^a \nabla_a u^b $, but the fact that $\zeta^a$ in general does not have good foliation properties ---  in contrast to $u^a$ --- makes the mathematics 
  inelegant and the interpretations rather contrived. I believe it is the use of $\zeta^a$ rather than $\xi^a$ in the past literature which has led to the missing of some of these results, which, as we will see, are mathematically rather simple.
  
  I use the (-- + + +) signature and, most of the time, units with $c=1,\hbar=1,k_B=1,16\pi G=1$, so that Einstein's equations reduce to $2G_{ab}=T_{ab}$. The Latin letters run through 0-3 while the Greek letters run through 1-3. I define (...) and [...] for (anti)symmetrization of tensor indices \textit{without} a factor $(1/2)$. I also use the convention that $\delta $ (something) $= - \pounds $ (something) with a \textit{relative minus sign} when the variation of (something) is produced by a diffeomorphism represented by Lie differentiation along a vector field.

  \section{Introducing the dramatis personae: A route to Gravity}\label{sec:2}

   Recent research \cite{KBP} motivated by the emergent paradigm of gravity shows that the description of classical gravity simplifies significantly if we use the variables $f^{ab}\equiv \sqrt{-g}g^{ab}$ and the corresponding canonical momentum $N^c_{ab}$. These `old variables' were studied \cite{a36,a37,a38} in the early days of general relativity (and were used sporadically later on in the literature, like,   e.g., in \cite{a45,a44}) but have not acquired the popularity they deserve. My first task  is to advertise the virtues of this description.

  Classical mechanics of a single degree of freedom $q(t)$ can be obtained from an action principle based on the Lagrangian $L_q\equiv p\dot q - H(p,q)$. Varying $p$ and $q$ independently, one sees that: (a) To get sensible equations of motion, we need $\delta q =0$ at the boundary while $\delta p $ is arbitrary. (b) When $\delta q=0$ at the boundary, the equations of motion are $\partial_t q = (\partial H/\partial p)$ and $\partial_t p = -(\partial H/\partial q)$. 
The subscript $q$ on $L_q$ is to remind us that $q$ is kept fixed at the end points to obtain the equations of motion. We can ask whether there is another Lagrangian which will lead to the \textit{same equations of motion} when $p$ is kept fixed. There indeed exists one, which is given by
\begin{equation}
L_p \equiv L_q - \frac{d(pq)}{dt }= - q \dot p - H(p,q)                         
\end{equation} 
 It is easy to see that if we vary $L_p$, treating again $q$ and $p$ as independent, we can get the same equation of motion with $\delta p =0$ at the boundary and $\delta q$ arbitrary. Note that $L_q=L_q(q,\dot q,p)$ while $L_p=L_p(q,p,\dot p)$.

We also know that we could have got the same equations of motion by treating $L_q$ as just a function of $q,\dot q$ and identifying  $p=p(q,\dot q)$ as $p=\partial L_q/\partial \dot q$ without treating it as an independent variable.  In such an approach, there is  a significant difference between $L_p$ and $L_q$. When $L_q=L_q(q,\dot q)$ depends  only on up to first derivatives of $q$, the $L_p=L_p(q,\dot q,\ddot q)$ will (in general) depend on second derivatives of $q$ through the term $d[q(\partial L/\partial \dot q)]/dt$. Normally, if a Lagrangian depends on the second derivatives of $q$, the equations of motion can be third order in $q$. But  \textit{with this specific kind of dependence}, the equations of motion will still be second order in $q$ when we keep $p$ fixed at the end points. Thus, it is easy\footnote{In fact, given  $L_q(q,\dot q)$ one can construct a Lagrangian which leads to the same equations of motion when some arbitrary function  $C(q,\dot q)$ is kept fixed at the boundary; see \cite{ayan} for details.} to construct, even in classical mechanics, a Lagrangian containing $\ddot q$, which only leads to second order equations of motion.\footnote{The Lagrangian $L_p=L_q-d(qp)/dt$ is almost never used in classical mechanics or field theory. For a free particle,  $L_p(q,\dot q,\ddot q)=-(1/2)\dot q^2-q\ddot q$; this should not be confused with   the Lagrangian $-(1/2)q (d^/dt^2)q$ which \textit{is} sometimes used.} 

What is not stressed in textbooks is that  we can do the same thing in field theory. The field equations, $\square \phi = - V'(\phi)$ for a scalar field $\phi$, say, can be obtained using  either of the two Lagrangians: 
\begin{equation}
L_\phi = p^a \partial_a \phi - H; \quad L_p = - \phi \partial_a p^a - H = L_\phi - \partial_a (\phi p^a); \quad H= (1/2) p^ap_a + V(\phi). 
\end{equation} 
These Lagrangians, again, differ by a total divergence. In  contrast to the text book description, we are here treating $H$ as a \textit{Lorentz scalar} and not as the time component of a four-vector; further, full Lorentz invariance is maintained without any (1+3) split. Note that, while  the $p^a\partial_a\phi$ term does not involve a metric, $p^ap_a=\eta_{ab}p^ap^b$ needs a background metric to `lower the index' on $p^a$.  We can also work out the dynamics of a vector field and, in particular, the  $U(1)$ gauge field in an analogous manner. It is straightforward to generalize the description to a curved spacetime by the usual prescriptions.
  
The situation becomes really interesting when we consider the theory of a symmetric second rank tensor field described by a matrix $f$ with elements denoted by $f^{ab}$. The ``momenta'' corresponding to $f^{ab}$ will be denoted by $N^c_{ab}$ which, in turn, can be thought of as elements of four matrices $N^c$. As in the previous cases, we can consider two possible Lagrangians (which differ from each other by a four-divergence) to describe the theory:
  \begin{equation}
 L_f = N^{c}_{ab}\partial_c f^{ab} - \mathcal{H}_g(f^{ij}, N^k_{lm}); \qquad
  L_N = -f^{ab}   \partial_cN^{c}_{ab}  - \mathcal{H}_g
  = L_f - \partial_c(f^{ab} N^{c}_{ab})
\label{lfdef}
\end{equation}
with a very specific choice:
\begin{equation}
\mathcal{H}_g=f^{ab}(N^{c}_{ad}N^{d}_{bc}- \frac{1}{3} N^{c}_{ac}N^{d}_{bd}) 
\label{Hg}
\end{equation}
Note that, unlike the scalar and vector cases, we do \textit{not} need to use $\g d^4x$ as integration measure; the action is obtained in \textit{any} coordinate system by integrating over just $d^4x=dx^0dx^1dx^2dx^3$. This makes the action a polynomial function of the basic variables $(f^{ab},N^i_{jk})$, unlike what happens when we use $g_{ab}$ due to the presence of the  $\g$ term. Further,
these Lagrangians are defined without the use of any background metric (in contrast to the scalar/vector field theories) and the contractions in \eq{lfdef} and \eq{Hg} are purely combinatoric operations involving only $\delta^i_j$. 
In fact, instead of thinking of $f^{ab}$ as a field in spacetime etc., we can think of 
these Lagrangians 
as describing the (abstract) dynamics of five $4\times 4$ matrices ($f, N^c$), the elements of each of which depend on four parameters $q^i$. The matrix action is determined by integrating the $L_f$ or $L_N$ over $d^4q$. The variational principle using $L_f$ (with $f$ fixed at the boundary) or with $L_N$ (with $N^c$ fixed at the boundary) will lead to the dynamical equations for the matrices. (This  fact allows a different way of approaching gravitational dynamics which I will comment on in Sec.~\ref{sec:2.2}.)

Incredibly enough, 
the \textit{resulting equations are identical to those of Einstein's gravity} (without sources; it can be easily extended to take care of matter sources \cite{KBP}) if we identify the arbitrary curved spacetime coordinates $x^i$ with the parameters  $q^i$ and set $f^{ab} \equiv \sqrt{-g} g^{ab}$. The equations then imply the further identification:
 \begin{equation} 
N^{a}_{bc} = -\Gamma^{a}_{bc}+\frac{1}{2}(\Gamma^d_{bd}\delta^{a}_{c}+\Gamma^d_{cd}\delta^{a}_{b}) = Q^{ad}_{be}\ \Gamma^e_{cd} + Q^{ad}_{ce}\ \Gamma^e_{bd}
\label{NGamma}                       
\end{equation}  
where $Q^{ab}_{cd} = (1/2) [\delta^a_c\delta^b_d - \delta^a_d\delta^b_c]$ is the determinant tensor.
(For a detailed proof, see \cite{KBP}.)
Thus, gravitational dynamics in an \textit{arbitrary} coordinate system with labels $x^i$ and metric functions $g_{ab}(x^i)$ can be obtained from the dynamics of five matrices ($f, N^c$).

Another surprise  is that the momentum space Lagrangian $L_N$ is numerically just $\g R$ where $R$ is the  curvature scalar!
 That is, treating $N^i_{jk}$ as a function of $f^{ab}$ given by \eq{NGamma}, one can show \cite{KBP} that 
\begin{equation}
L_N = L_f - \partial_c (f^{ab}N^c_{ab}) = - f^{ab} \partial_c N^c_{ab} - \mathcal{H}_g
= \g R                                              
\end{equation} 
where $R$ is the standard curvature scalar of differential geometry.
The reason  the Einstein-Hilbert action  leads to second order equations is now clear: It is a momentum space Lagrangian in terms of  the  variables $(f,N^a)$.

As we shall see, the variables $(f,N^a)$ will play an extremely important role in the rest of the paper. In view of this fact, I will next describe several important properties of these variables and connect them up with the standard description to enhance familiarity. (For more details see, \cite{KBP}.)

\subsection{Description of gravity in terms of $(f^{ab},N^c_{ab})$}\label{sec:2.1}

To begin with, the standard Einstein-Hilbert Lagrangian
(in units $16\pi G=1$)
\begin{equation}
 A_{\rm EH}= \int_\mathcal{V} d^4x\, \mathcal{L}_{\rm EH} =\int_\mathcal{V} d^4x\sqrt{-g}\, R = \int_\mathcal{V} d^4x\, \g Q^{ab}_{cd} R^{cd}_{ab}
\label{A_EH}
\end{equation}
can be decomposed into a quadratic Lagrangian (called  the $\Gamma^2$ Lagrangian or the  Einstein-Schrodinger Lagrangian) and a divergence term as:
$ \mathcal{L}_{\rm EH} = \mathcal{L}_{\rm quad}+\mathcal{L}_{\rm sur}$,
with
\begin{equation}
\mathcal{L}_{\rm quad}\equiv2 \sqrt{-g}Q_a^{\phantom{a}bcd}\Gamma^a_{dk}\Gamma^k_{bc};\qquad
\mathcal{L}_{\rm sur}\equiv2\partial_c\left[\sqrt{-g}Q_a^{\phantom{a}bcd}\Gamma^a_{bd}\right].
\label{bulksur}
\end{equation}
These expressions take simpler forms when we use the variables $f^{ab}$ and $N^a_{bc}$ treating $N^a_{bc}$ as a specified function of $f^{ab}$. We again have $\mathcal{L}_{EH}=\mathcal{L}_{\rm quad}+\mathcal{L}_{\rm sur}$ with
\begin{equation}
 \mathcal{L}_{\rm quad}=\frac{1}{2}N^a_{bc}\partial_af^{bc}; \quad
\mathcal{L}_{\rm sur}=-\partial_c(f^{ab}N_{ab}^c)
\end{equation} 
The Ricci tensor can be expressed in terms of $N^a_{bc}$ as:
\begin{equation}
 R_{ab}=-(\partial_c N^{c}_{ab}+N^{c}_{ad}N^{d}_{bc}- \frac{1}{3} N^{c}_{ac}N^{d}_{bd})
\end{equation} 
Though $N^a_{bc}$ is not a tensor, its variation $\delta N^a_{bc}$ is a tensor which can be related to the variation of the  Ricci tensor, $\delta R_{ab}$, again, in a remarkably simple form:
\begin{equation}
 \delta R_{ab} = - \nabla_c (\delta N^c_{ab})
 \label{rabfromn}
\end{equation}
(Incidentally the result in \eq{rabfromn} allows us to formulate  the gravitational field equations in a novel manner which I have commented upon in Sec.~\ref{sec:2.2}). 
As a corollary, when the variations are induced by a Lie derivative along an arbitrary vector field $q^a$, \eq{rabfromn} becomes
\begin{equation}
 \pounds_q R_{ab} = - \nabla_c (\pounds_q N^c_{ab})
 \label{rabfromLie}
\end{equation}
The Lie derivative of  non-tensorial objects like $\Gamma^a_{bc}$ can be defined from first principles and the explicit form of $\pounds_q \Gamma^a_{bc}$ is given later in \eq{lvgamma}; the Lie derivative of  $N^a_{bc}$ follows on using \eq{NGamma}. Since $N^a_{bc}$ is the momentum density, \eq{rabfromLie} shows that the Ricci tensor is closely related to the variation of gravitational momentum when Lie transported along any vector field.

Multiplying \eq{rabfromn} by $f^{ab}=\g g^{ab}$ we immediately get: 
\begin{equation}
 f^{ab} \df R_{ab} =\delta(\sqrt{-g}R)-R_{ab}\delta f^{ab}=-\partial_{c}[f^{ik}\delta N^{c}_{ik}],
\end{equation} 
which, in turn, gives the  variation of the Einstein-Hilbert action in just one step:
\begin{equation}
 \delta(\sqrt{-g}R)=  R_{ab}\delta f^{ab}-\partial_{c}[f^{ik}\delta N^{c}_{ik}]
 =\g[G_{ab}\delta g^{ab}-\nabla_{c}(g^{ik}\delta N^{c}_{ik})]
 \label{varRNf}
\end{equation} 
To get the second equality, we have used the fact that for any two index object $X_{ab}$, we have $X_{ab}\delta f^{ab}=(X_{ab}-(1/2)g_{ab}X)\delta g^{ab}$. If the variations in \eq{varRNf} arise due to a Lie differentiation along a vector field $q$, then on multiplying \eq{rabfromLie} by $f^{ab}=\g g^{ab}$,  identical algebra gives [on using $\pounds_q(\sqrt{-g}R)=\g\nabla_a(Rq^a),  G_{ab}\pounds_q g^{ab}=-2\nabla_a(G^a_b q^b)$] the conservation law $\nabla_aJ^a=0$ for the current:
\begin{equation}
 J^a[q]=2G^a_b q^b+Rq^a+g^{ik}\pounds_q N^{a}_{ik}=2R^a_b q^b+g^{ik}\pounds_q N^{a}_{ik}
 \label{ja}
\end{equation} 
This current will again play a vital role in our discussion throughout the paper.

 Further, we can define a vector $P^a[q]$ associated with any vector field $q^a$ by rewriting \eq{ja} as
\begin{equation}
P^a[q]\equiv J^a[q]-2G^a_b q^b=g^{ik}\pounds_q N^{a}_{ik}+L_Hq^a
\label{defpa1}
\end{equation}
where $L_H=R$ is the Einstein-Hilbert Lagrangian. This vector is same as $J^a$ on-shell, when the equation of motion $G^a_b=0$ is satisfied and is conserved on-shell.
To see the additional significance of this vector, note that the Hamiltonian for gravity we started with in \eq{lfdef} has the form:
\begin{equation}
\mathcal{H}_g= -\g[ g^{ab} \partial_cN^{c}_{ab}+L_H] 
\end{equation} 
where we have used $L_N=R\g=L_H\g$. This, of course, is not a scalar density because of the $\partial_c$ in the expression. But the form of $\mathcal{H}_g$   suggests the replacement of $g^{ab}\partial_cN^{c}_{ab} $ by $g^{ab}\pounds_q N^{c}_{ab}$ along any vector field to obtain a scalar density for this term. To make the indices match, we replace $L_H$ by $L_Hq^a$. This is precisely the vector $P^a[q]$, with the sign chosen for future convenience.\footnote{With our choice of signature, in flat spacetime, say, $u_a=-\delta_a^0$ so that $u^a=\delta^a_0$ is future pointing. This means that in the dot products $u_av^a=-v^0$, we pick up the time component of $v^a$ with an extra minus sign rather than just pick up $v^0$ (which would have happened if we were using the opposite signature). This is not of concern most of the time and we will not worry about it. Occasionally, our sign choices are decided by this fact.}

In particular, if we chose $q^a=\xi^a\equiv Nu^a$ where $u_a$ is the normal to $t=$ constant surfaces, we can interpret $P^a[\xi]$ as the  gravitational four-momentum flux. When the equations of motion hold $2G^a_b \xi^b=T^a_b \xi^b$ can be thought of as the corresponding four-momentum flux of matter. The definition in \eq{defpa1}  tells us that:
\begin{equation}
J^a[\xi]=  P^a[\xi] +2G^a_b \xi^b
=P^a[\xi]+ T^a_b \xi^b\equiv P^a_{total}[\xi]                                                                                                                                                                                                                                                                                                                                                                                                                                                                                                                                                                                                                                                                                                                                                                \end{equation} 
allowing us to interpret the conserved current $J^a[\xi]$ as  the \textit{total} four-momentum flux. This interpretation will be made more precise and strengthened in Sec.~\ref{sec:5}.

These two results in \eq{varRNf} and \eq{ja} and the interpretation of \eq{defpa1} highlight the importance of the combination $g^{ik}\pounds_q N^{a}_{ik}$ which, as we saw, can be related to gravitational momentum flow.
We also know from \eq{varRNf} that $g^{ik}\delta N^{a}_{ik}$ is essentially the surface term in the variation of the  gravitational action:
\begin{equation}
(16\pi)\delta A_{\rm sur}=-\int_\mathcal{V} d^4x \partial_c [f^{ab}\delta N^c_{ab}]
=-\int_{\partial \mathcal{V}} d^3x \sqrt{h} n_c g^{ab}\delta N^c_{ab} 
\end{equation} 
It turns out that this variation can be given a simple thermodynamic interpretation. Consider a null surface with temperature $T=\kappa/2\pi$ and entropy density $s=\sqrt{\sigma}/4$ attributed to it by local Rindler observers who perceive it as a horizon with
  $\kappa$ being the surface gravity defined using a suitable null congruence and $\sqrt{\sigma}$ being the area element of the local horizon surface:
Then,   (see \cite{KBP} for details) the following results hold:

\begin{itemize}
 \item The boundary term in the action, evaluated over a null surface, can be interpreted \cite{a41} in terms of its heat content $Ts$ (or enthalpy density; see the footnote on page \pageref{pagerefone}); that is:
 \begin{equation}
  \frac{1}{16 \pi}\int d^3 \Sigma_c (N^{c}_{ab} f^{ab}) = \int d\lambda\ d^2x \left(\frac{\kappa}{2 \pi}\right)\left(\frac{\sqrt{\sigma}}{4}\right)=\int d\lambda\ d^2x\ T  s
 \end{equation} 
 
 \item More remarkably, the variations $f\delta N$ and $N\delta f$ have corresponding thermodynamic interpretations for a class of variations which preserve the null surface:
 \begin{eqnarray}
\frac{1}{16 \pi}\int  d^3 \Sigma_c(N^{c}_{ab}\df f^{ab})&=& \int d\lambda\ d^2x\ \left(\frac{\kappa}{2 \pi}\right)\df\left(\frac{\sqrt{\sigma}}{4}\right)= \int d\lambda\ d^2x\ T \df s; \\
\frac{1}{16 \pi}\int  d^3 \Sigma_c (f^{ab}\df N^{c}_{ab})&=& \int d\lambda\ d^2x\ \left(\frac{\sqrt{\sigma}}{4}\right)\df\left(\frac{\kappa}{2 \pi}\right)= \int d\lambda\ d^2x\ s \df T
\label{stSdT}
\end{eqnarray}
\end{itemize}
We see that ($f,N^c$) are not only \textit{dynamically conjugate} variables but their variations ($\delta f,\delta N^c$) exhibit \textit{thermodynamic conjugacy} in terms of corresponding  variations in $(\delta T, \delta S)$. Of these, we can think of $f$ as an extensive variable and $N^a$ as intensive variables, just as in conventional thermodynamics. Later discussion will fortify this idea.

If we consider the gravitational action principle in a region of spacetime bound by null surfaces  (the `causal diamond'), then the boundary condition on the surface corresponds to isothermality ($\delta T=0$) with respect to the local Rindler observers who perceive the null surfaces as  local Rindler horizons. It is obvious how thermodynamic considerations enhance our understanding of standard gravitational dynamics. We will see many more such examples in the sequel.

\subsection{Aside: Some additional comments}\label{sec:2.2}

There are a few comments regarding this formalism which I would like to mention here as an aside, though it is not relevant directly to the rest of the paper. Those interested in the main theme can directly proceed to Sec.~\ref{sec:3}.

First, in the standard approach, we use $g_{ab}$ rather than $f^{ab}\equiv\g g^{ab}$ as the dynamical variables and one may wonder whether similar results can be obtained with $g_{ab}$. This requires using the action:
\begin{equation}
A=\int d^4x \g L;\quad L=M^{cab}\partial_cg_{ab}-H 
\end{equation}
where
\begin{equation}
M^{abc}=g^{bd}g^{ce} \Gamma^{a}_{de} - \frac{1}{2}g^{bd} g^{ac} \Gamma^{e}_{de} - \frac{1}{2} g^{cd} g^{ab} \Gamma^{e}_{de}  - \frac{1}{2}g^{bc}V^{a}. \label{M_in_GammaV}
\end{equation}
with
\begin{equation}
V^{c}\equiv -g_{ab} M^{cab}=g^{ik} \Gamma^{c}_{ik}-g^{ck}\Gamma^{m}_{km}=2 Q_{a}^{\phantom{a}bcd}\Gamma^{a}_{bd}=2 Q^{ijkc}\partial_i g_{jk}=-\frac{1}{g}\partial_b(g g^{bc}).
 \label{Vc}
\end{equation}
and
\begin{equation}
 H =g^{bd}M^{c}_{dk}M^{k}_{bc}-\frac{2}{3}g^{ir}M^{k}_{ir}M^{d}_{dk}-\frac{1}{3}g^{bk}M^{i}_{bi}M^{d}_{dk}+\frac{1}{6}g_{dk}g^{ir}M^{d}_{ir}g^{xy}M^{k}_{xy} 
\end{equation} 
where $M^c_{ab}\equiv g_{ai}g_{bj}M^{cij}$ etc.  We see that: (i) We now have to define the action with a $\g d^4x$ measure rather than with just $d^4x$ introducing a  factor which is \textit{non-polynomial} in the dynamical variables unlike in the case of $(f^{ab},N^a_{bc})$.  (ii) The expressions are a lot more complicated compared to  when we use $(f^{ab},N^a_{bc})$ as our variables. We will see on several future occasions that the use of the $(f^{ab},N^a_{bc})$ pair simplifies the expressions. 

Second, the matrix dynamics in terms of the five $4 \times 4$ matrices $(f,N^a)$ achieves ``general covariance without general covariance'' in the following sense.
Consider a $4\times4$ matrix with elements denoted by $f^{ab}$ and another set of four $4\times4$
matrices [$N^0_{bc},N^1_{bc},N^2_{bc},N^3_{bc}$] with elements denoted by $N^a_{bc}$. We assume that all these elements $[f^{ab}(q^i),N^a_{bc}(q^i)]$  depend on four parameters $q^i$. We can now construct the index-free function $L_f(q^i)$ using \eq{lfdef} and \eq{Hg}. This does not require any metric and the index placements take care of summation with just Kronecker deltas.
(However, the definition of $\mathcal{H}_g$ is not a pure matrix operation but involves the components.)
We can integrate $L_f(q^i)$ over a range of parameters $q^i$ with the Cartesian measure $d^4q$ to define an abstract matrix action. The extremisation of this action when the matrices $(f,N^a)$ are varied, will lead to a set of equations which determine the relation between $\partial_c f$ and $N^a$ as well as their dependence on $q^i$.
Now consider the real spacetime manifold described by  the metric functions $g_{ab}(x^i)$ in some specific coordinate system. If we identify the coordinates $x^i$ with our parameters $q^i$ and the combination $\g g^{ab}$ with $f^{ab}$, then the matrix equations we obtained determine the spacetime metric functions in these coordinates. If we now choose another set of coordinates in the spacetime, the metric functions will change. \textit{But we do not have to change anything as regards the abstract matrix description; in particular we do not have to multiply $d^4q$ by a Jacobian.} We simply identify the new coordinates with $q^i$ and  the new  $\g g^{ab}$ with $f^{ab}$ and  everything will work out quite trivially. This separation of kinematics (coordinates, explicit form of metric functions, their change under coordinate transformations, etc.) from the dynamics (described once and for all by the equations satisfied by  the matrices $(f,N^a)$) is an attractive conceptual feature of this approach. 

Third, one can easily introduce more `pre-geometric' variables $\lambda_A$, say, and treat the matrix element $f^{ab}(q^i)$ as some kind of coarse-grained object after averaging over $\lambda_A$, like
$f^{ab}(q^i)=\langle f^{ab}(q^i;\lambda_A)\rangle$ where the averaging is over the pre-geometric variables $\lambda_A$ with some measure. This shows that the current approach may be useful for introducing and investigating pre-geometric structures.

Fourth, the nature of gravity is completely contained in the form of the Hamiltonian $ \mathcal{H}_g$. On general grounds, one would expect it to be a quadratic in $N^i_{jk}$. The choice of the specific form of the quadratic function depends on the symmetries we want to impose on the theory. 
 There is a natural, covariant, separation of any metric into a conformal form $g_{ab}=\Omega^2\bar{g}_{ab}$ with $\bar{g}_{ab}$ being a metric of unit determinant. This allows reformulation of the theory in terms of the variables $(\Omega,\bar{f}^{ab})$, which has some interesting features that I hope to discuss in a future publication.  

Fifth, the relation \eq{rabfromn} allows a `variational' formulation of gravitational dynamics.
Suppose we have a source $\mathscr{F}_{ab}(s)\equiv(T_{ab}-(1/2)g_{ab}T)$ where $T_{ab}(s)$ is the stress-tensor which depends on some parameter $s$ (mass, charge, cosmological constant, anything....). For every value of $s$, we can, in principle, solve the field equations and obtain $g_{ab}(s),f_{ab}(s),N^c_{ab}(s),....$ etc. Let an overdot denote the derivative of any of these with respect to $s$, like $\overset{\bm{\centerdot}}{\mathscr{F}}_{ab}=d \mathscr{F}_{ab}/ds$, etc. Then the gravitational field equations can be written simply as:
\begin{equation}
  \nabla_c (\overset{\bm{\centerdot}}{N}{}^c_{ab})=-8\pi \overset{\bm{\centerdot}}{\mathscr{F}}_{ab}
\label{dotn}
\end{equation} 
which tells us how the geometry \textit{changes} when we \textit{change} the source. Given $ \overset{\bm{\centerdot}}{\mathscr{F}}_{ab}$, we can solve \eq{dotn} for $\overset{\bm{\centerdot}}{N}{}^a_{bc}(s)$, and find $N^a_{bc}(s)$ by integrating $\overset{\bm{\centerdot}}{N}{}^a_{bc}(s)$ over $s$ with flat spacetime as  the initial condition, say. We again see, from \eq{dotn},  the importance of 
$N^a_{bc}$; the field equations directly determine how it changes when the source changes. Such a description relating \textit{changes} in the geometry to \textit{changes} in the source may be closer in spirit to thermodynamics.
These  aspects open up interesting avenues for further work.

\section{Conserved currents from identities in differential geometry}\label{sec:3}

In the later sections, I will extensively use certain  conserved currents (usually called Noether currents) to describe the gravitational dynamics. 
Conventionally, there is a tendency in literature to link the existence of such conserved currents with the diffeomorphism invariance of an action.
If one follows the route of  deriving the conserved currents from the invariance properties of a given action functional, then it is somewhat of a circuitous reasoning --- though interesting --- to redescribe dynamics in terms of these currents. (After all, complete dynamics can be obtained if  the form of the action  is known.) Fortunately, this approach is unnecessary and --- to a great extent --- complicates matters. 
I will  show how the variables ($f, N^a$) are closely related to conserved (Noether) currents for  vector fields purely because of some ---in fact,  rather trivial --- identities in differential geometry. \textit{This delinks the existence of conserved (Noether) currents from the invariance properties of any action principle or such dynamical considerations.}

 There are (at least) two ways of obtaining the standard Noether current associated with a vector field $v^a$ \textit{without ever mentioning the action principle for gravity or its diffeomorphism invariance!}. These involve  just  (i) using  either the antisymmetric part of $\nabla_iv_j$ and an identity 
for $R^i_jv^j$ or (ii)   contracting the Lie derivative of $\Gamma$ suitably. This shows that
the resulting conservation law $\nabla_aJ^a=0$,  for example, is just a differential geometric identity and \textit{has nothing to do  with the gravitational dynamics}. Delinking the \textit{form} of  the Noether current from any action principle allows us to reinterpret gravitational dynamics later on using Noether currents, without introducing any circuitous reasoning. 

We begin from the elementary fact that the derivative $\nabla_k v_j$ of  any vector field $v^j$  can be decomposed into the anti-symmetric and symmetric parts by 
\begin{equation}
\nabla^j v^k + \nabla^k v^j \equiv \nabla^{(j} v^{k)} \equiv S^{jk};\qquad
 \nabla^j v^k - \nabla^k v^j \equiv \nabla^{[j} v^{k]} \equiv J^{jk}
\end{equation}
[Recall that I define (...) and [...]  \textit{without} a factor $(1/2)$.]   The antisymmetric part $J^{lm}$  leads to a conserved current $J^i\equiv \nabla_k J^{ik}$; in other words, from every vector field $v^k$ in the spacetime we can obtain a conserved current, fairly trivially.\footnote{In fact, the converse is also true; given a conserved current $J^i$ one can always find the corresponding vector field $v^k$. Since $J^{il}$ is invariant under the ``gauge transformation'' $v^k \to v^k + \partial^k f$, we in fact have an infinite number of such vectors associated with a given conserved current. This is just electromagnetism in disguise.} 

To find an explicit form for this current and explore its connection with $N^a_{bc}$, we proceed as follows:
Substituting the decomposition,
$ \nabla^j v^k = (1/2) \left( J^{jk} + S^{jk}\right)$, 
in the standard identity
\begin{equation}
 \nabla_k \left( \nabla^j v^k\right) - \nabla^j \left( \nabla_k v^k\right) = R^j_l v^l
\label{ddv}
\end{equation} 
we get:
\begin{equation}
 \nabla_k J^{jk} + \nabla_k ( S^{jk} - g^{jk} S ) = 2 R^j_l v^l
\label{1.5}
\end{equation} 
The first term in the above equation is the conserved current $J^k$, while the physical meaning of  the second term can be obtained by noting that
the symmetric part $S^{jk}$ gives the change in the metric tensor $g^{jk}$ under the diffeomorphism $x^a \to x^a + v^a$. That is:
\begin{equation}
 \pounds_v g^{ab}=-[ \nabla^a v^b + \nabla^b v^a]= -S^{ab}\equiv - \delta g^{ab} 
\end{equation} 
(Recall our convention that $\delta = - \pounds $).
From the definition in \eq{NGamma} we can also obtain the explicit result:
\begin{equation}
 g^{ab}\delta N^c_{ab}  =\nabla_b(\delta g^{cb}-g^{cb}g_{ik}\delta g^{ik})
\label{secterm} 
\end{equation}
which is valid for arbitrary variations, not just diffeomorphism  (see, e.g., p. 549 of Ref.\cite{tpgravity}). In the case of diffeomorphisms (with $\delta N^c_{ab}=-\pounds_v N^c_{ab},\delta g^{ab} = S^{ab}$), we get:
\begin{equation}
 g^{ab}\pounds_v N^c_{ab}\equiv B^c[v]=-\nabla_b(S^{bc}-g^{cb}S);\quad
 \nabla_c (g^{ab}\pounds_v N^c_{ab})\equiv\nabla_cB^c
\end{equation} 
 Comparing this with the form of the  second term on the left hand side of \eq{1.5}, we see that the second term is just $(-g^{bc} \pounds_v N^j_{bc})$. 
This gives us the explicit form of the conserved current
\begin{equation}
 J^a[v] = \nabla_b J^{ab} [v] = 2 R^a_b v^b + g^{ij} \pounds_v N^a_{ij}
=2 R^a_b v^b +B^a[v]
\label{noe}
\end{equation} 
which, of course, is the same one as obtained before in \eq{ja}, but now we see that its conservation follows trivially from the antisymmetry of $\nabla^{[j} v^{k]}$ which was far from obvious in the derivation leading to \eq{ja}.

An alternate, faster,  way of obtaining the same result is as follows: From the Lie derivative of the connection
\begin{equation}
\pounds_v\Gamma^a_{bc}=\nabla_b \nabla_c v^a+R^a_{\phantom{a}cmb}v^m
\label{lvgamma}
\end{equation} 
one can obtain, on using \eq{NGamma}, the relation:
\begin{equation}
2Q^{adc}_{\phantom{adc}e}\pounds_v\Gamma^e_{cd}=
g^{bc}\pounds_vN^a_{bc} =\nabla_bJ^{ab}-2R^a_bv^b
\end{equation} 
which is the same as the one obtained earlier. 

We thus see that there are two routes to obtaining the form of the Noether current: (a) from the antisymmetric part of $\nabla_iv_j$ using the  identity 
in \eq{ddv} or (b) by contracting the Lie derivative of $\Gamma$ suitably.
In either procedure, the natural object we obtain is $J^{ab}$ (called the  Noether potential in general or  the Komar superpotential in this particular case, which are just other names for $\nabla^{[j} v^{k]}$!) and $J^a$ is \textit{derived} from it. \textit{This route is unique and natural}. In contrast, in the conventional approach one \textit{first} obtains the Noether current $J^a$ and \textit{then} looks for a Noether potential $J^{ab}$ thereby agonizing over its non-uniqueness.

Thus, with every vector field $v^j$ we can associate a conserved current $J^i[v]$ given by  \eq{noe}. One could think of this as a trivial algebraic result arising from the antisymmetry of $J^{ij}$ related to the vector field. On the other hand,  it also \textit{happens to be } the Noether current arising from the diffeomorphism invariance of the scalar $R$. Nobody knows of a clear \textit{physical} reason as to why this happens; in other words, nobody knows how to guess this answer without doing the calculation. This fact, probably, has some physical significance and --- in the discussions later --- I will try to emphasize this to a certain extent.

\section{Bulk gravitational dynamics and surface thermodynamics}\label{sec:4}

For the rest of the paper, we will consider a spacetime  foliated by a series of space-like hypersurfaces defined by constant values for a suitable scalar field $t(x)$. We define the unit normal to $t=$ constant surfaces by 
$u_a  = - N \nabla_a t$; if we choose our time coordinate to coincide with the hypersurface label, then   $u_a = - N\delta_a^0$.  We next define a `time-development' vector $\zeta^a$ by the  invariant condition $\zeta^a \nabla_a t =1$.  (In the preferred coordinate system, we can choose  $\zeta^a = \delta^a_0$).  In general, $\zeta^a$ and $u^a$ will not be in the same direction and we have
$\zeta^a = - (\zeta^b u_b) u^a + N^a$
 where $u_aN^a=0$ and $N^a=h^a_b\zeta^b$ where $h^a_b\equiv \delta^a_b+u^au_b$ is the projection tensor. In component form,
 $\zeta^a=Nu^a+N^a$ with
 $N^a=(0,N^\alpha)$. 

This decomposition also introduces the vector $\xi^a\equiv -(\zeta^b u_b) u^a= Nu^a$ which will turn out to be of considerable importance in what follows.  When we  impose the coordinate condition such that $g_{0\alpha} =0$ in a local region,  we will have $\zeta^a =  \xi^a$ and, if the spacetime is static, we can identify $\xi^a$ with the time-like Killing vector.
Thus, in \textit{any} spacetime, there exist two \textit{natural} diffeomorphisms, with vector fields  
\begin{equation}
\xi^a= Nu^a =-N^2 \nabla_a t; \quad \zeta^a=Nu^a+N^a                                                    \end{equation}  
The latter one $\zeta^a$  has attracted a fair amount of attention in the past  literature. I will, however, show that as far as Noether currents and related dynamics go, the vector $\xi^a$ possesses more interesting properties. So, I will begin with a discussion of these properties and later comment on corresponding results \textit{vis-a-vis} the vector field $\zeta^a$ in Sec.~\ref{sec:7.3}. (Incidentally, the Noether current for $u^a$  is also \textit{not} interesting compared to the one for $\xi^a$; see the discussion in Appendix \ref{appen9.2}.)

\subsection{Noether charge as the surface heat content}\label{sec:4.1}

We will begin by computing the  Noether current for $\xi^a$. To calculate this current, it is convenient to use  an identity connecting Noether currents for two vector fields $q^a$ and $v^a \equiv f(x) q^a$. It can be shown that [see Appendix \ref{appen9.1}; \eq{genresult1}]
\begin{equation}
 q_a J^a [fq] - f q_a J^a [q] = \nabla_b\left[ \left( q^a q^b - q^2 g^{ab} \right) \nabla_a f\right]
 \label{genresult}
\end{equation} 
which is particularly useful if $q_a=\nabla_a\phi$ so that $J^{ab}[q]=0$.
If we   use \eq{genresult} with $q_a = - \xi_a/N^2,\ f= - N^2$, we can obtain 
a nice result for the Noether charge density [see Appendix \ref{appen9.2}; \eq{fin1A}]:
\begin{equation}
u_aJ^a(\xi)=2D_\alpha (Na^\alpha)
\label{fin1}
\end{equation} 
where $a^i\equiv u^j\nabla_ju^i$ is the acceleration and  $D_ia^i= D_\alpha a^\alpha =  \nabla_i a^i - a^2$ where $D_i$ is the covariant derivative on the $t=$ constant surface. 
The acceleration $a_i$, which will occur repeatedly in our discussion, has the explicit form
[see Appendix \ref{appen9.2}; \eq{adef1}]
\begin{equation}
 Na_i=Nu^l\nabla_lu_i=h^j_i\nabla_jN.
\end{equation}
The combination  $Na_i=\xi^l\nabla_lu_i$ measures the change in velocity along our chosen time-development vector $\xi^l$.

Integrating \eq{fin1}  over $\sqrt{h}d^3x$ to obtain the total Noether charge, we find that the flux of the acceleration is essentially the total Noether charge contained inside a volume. Noting that we have set $16\pi G=1$ and adding  the correct proportionality constant (now with $G=L_P^2!$), we get:
 \begin{equation}
\int_\mathcal{V}\sqrt{h}\,d^3x\ u_aJ^a[\xi]=\int_\mathcal{V}d\Sigma_aJ^a[\xi]=
\int_{\partial\mathcal{V}}\frac{\sqrt{\sigma}\, d^2x}{8\pi L_P^2} (Nr_\alpha a^\alpha)
\label{flux2}
\end{equation} 
This result is valid 
for any region $\mathcal{V}$ in any spacetime.
Let us now choose the boundary to be a $N(t,\mathbf{x})$= constant surface within the $t=$ constant surface.
In the above expression, $r_\alpha$ is then the  normal to  the $N(t,\mathbf{x})$= constant surface within the $t=$ constant surface. Therefore, one can write $r_\alpha\propto D_\alpha N$ or as
$r_i\propto h^j_i\nabla_jN$ where $h^i_j=\delta^i_j+u^iu_j$ is the projection tensor to the  $t=$ constant surface.  Since $Na_i=h^j_i\nabla_jN$, it follows that $r_i$ and $a_i$ are in the same direction even in the most general case (non-static, $N_\alpha\neq0$). Normalizing it, we can take $r_\alpha=  \epsilon a_\alpha/a$ where $a$ is the magnitude of the acceleration, ensuring that $r_\alpha$ is always outward pointing.
 This leads to $Nr_\alpha a^\alpha=  \epsilon Na=  \epsilon (h^{ij}\nabla_iN\nabla_jN)^{1/2}$.
 So, if we choose the boundary to be a surface with $N$=constant (which is a generalization of the notion of an equipotential surface), we can interpret $T_{loc}=Na/2\pi$ as the (Tolman redshifted) local Davies-Unruh temperature of the observers with four-velocity $u_a=-N\delta_a^0$. These observers, who are moving normal to the $t=$constant hypersurfaces will have the acceleration $a$ with respect to the local freely falling observers. The local vacuum of the freely falling frame will appear to be a thermal state with temperature $T_{loc}=Na/2\pi$ to these observers. So we can write:
\begin{equation}
2\int_\mathcal{V}\sqrt{h}\, d^3x\ u_aJ^a[\xi]=  \epsilon
\int_{\partial\mathcal{V}}\frac{\sqrt{\sigma}\, d^2x}{L_P^2} \left(\frac{1}{2}T_{loc}\right)
\label{ib1}
\end{equation}
Thus, (twice) the Noether charge contained in a $N$= constant surface is equal to the  equipartition energy of the surface when we attribute $(1/L_P^2)$ degrees of freedom per unit area.  
(We will comment on the factor 2 in a moment.)

Another, equivalent interpretation emerges, if we think of
$s=\sqrt{\sigma}/4L_P^2$ as the analogue of the entropy density. Then we get, directly from \eq{flux2}, the result: 
\begin{equation}
\int_\mathcal{V}\sqrt{h}d^3x u_aJ^a[\xi]
=  \epsilon \int_{\partial\mathcal{V}}d^2x\ Ts
\label{flux3}
\end{equation}
which is the heat (enthalpy)  density ($TS/A$) of the boundary surface. This interpretation of a quarter of the area as entropy is obvious in those limiting cases in which the boundary surface becomes a horizon (like e.g., the surface $t=$ constant, $r=2M+\epsilon$ in the Schwarzschild metric in the limit of $\epsilon\to0$). But one can define this notion in a more general context, in terms of local Rindler observers, as follows: Any sufficiently small patch of the boundary can be thought of as a small section of the $y-z$ plane which acts as a cross-section of the local Rindler horizon for a suitably defined observer accelerating along the $x-$ axis with acceleration $a$. Then, one can again attribute the entropy density $s=\sqrt{\sigma}/4$ and the temperature $T_{loc}=Na/2\pi$ to this patch of area from the point of view of the local Rindler observer. 

Let me  comment on the factor 2 on the left hand side of \eq{ib1} and connect it to a result familiar to general relativists (also see \cite{a40}). The integral on the right of \eq{ib1} gives $(1/2)TA=2TS$ if we take (for the sake of illustration) $T$= constant on the boundary and $S=A/4$. Therefore, the Noether charge $Q$ is just the heat content (enthalpy) $Q=TS$, which is also clear from \eq{flux3}. Thus, the Noether charge is \textit{half} of the thermal, equipartition, energy of the surface $(1/2)TA=2TS$ if we attribute $(1/2)T$ per surface degree of freedom. In the case of the Schwarzschild geometry, say, the  thermal, equipartition, energy of the surface is just the total mass $M=2TS$. \textit{But what the Noether charge measures is the heat content (enthalpy) $E-F=TS$ which is precisely $(M/2)$. } In fact, this is a well-known ``problem'' (see, e.g. \cite{mby2}) when one tries to define total mass of a spacetime (which asymptotically tends to the Schwarzschild limit) using the so-called Komar integral. In this context, $\xi^a$ will become the standard timelike Killing vector and the Noether potential will be the Komar potential. The integral one performs with the Killing vector $\xi^a$ is identical to the computation of the Noether charge above and one gets $(M/2)$. In classical relativity, this was considered very puzzling because in classical general relativity we (at best!) only have a notion of energy but no notion of heat content ($TS$), free energy ($F=E-TS$), etc. The thermodynamic perspective --- which requires $\hbar$ to define the Davies-Unruh temperature $k_BT=(\hbar/c)(\kappa/2\pi)$) from an acceleration $\kappa$ --- tells us that the Noether charge is the heat content (enthalpy) $TS$ and \textit{not} the energy $2TS$, and that the result \textit{must} be $M/2$ for consistency.
In short, classical general relativity can only interpret $M$ physically (as energy) while the thermodynamic considerations allow us to \textit{also} interpret $M/2$ physically as the heat content $TS$. \textit{This is yet another case of thermodynamic considerations throwing light on some puzzling features of classical general relativity.} 

Getting back to our main theme, we see that the Noether charge has a delightfully simple interpretation as the surface heat content (or half the surface equipartition energy) in the most general context. I did not have to assume static nature, existence of Killing vectors, asymptotic behaviour, etc. As long as the boundary is an $N(t,\mathbf{x})=$ constant surface in an otherwise general context, the Noether charge for the time-development vector, contained in a bulk region of space is equal to the boundary enthalpy. To do this,
 we have only used results from quantum field theory in curved spacetime and have \textit{not} used the gravitational field equations. 

I will now show how the gravitational field equations themselves acquire an interesting interpretation in this language.

\subsection{Gravitational dynamics and holographic equipartition}\label{sec:4.2}

Since the Noether charge is  related to $TS$ while $\delta N^a_{bc} $ is related to $\delta T$, it is obvious that one can interpret the gravitational dynamics in terms of thermodynamic variables using Noether currents. To do this, we 
take the dot product of the Noether current $J^a[\xi]$ (given in \eq{noe} with $v^a=\xi^a$)
with $u_a$ and use the result in \eq{fin1} to obtain:
\begin{equation}
 u_a g^{ij} \pounds_\xi N^a_{ij} = D_\alpha ( 2 N a^\alpha) - 2 N R_{ab} u^a u^b
\label{jforxi}
\end{equation} 
We next integrate this result over a 3-dimensional region $\mathcal{R}$ with the measure $\sqrt{h}\, d^3x$, resulting in:
\begin{equation}
 \int_\mathcal{R} d^3 x \, \sqrt{h}\, u_a\,  g^{ij}  (\pounds_\xi N^a_{ij})
= \int_{\partial\mathcal{R}} d^2 x \, \sqrt{\sigma}\,r_\alpha (2Na^\alpha) - \int_\mathcal{R} d^3 x \, \sqrt{h}\, (2Nu^au^b R_{ab})
\end{equation} 
where $r_\alpha$ is the normal to the boundary of the 3-dimensional region. This result, again, has a simple physical meaning if we choose the boundary to be a $N(t,\mathbf{x})$= constant surface within the $t=$ constant surface.  We introduce the gravitational dynamics through $R_{ab} = (8\pi L_P^2)\mathscr{F}_{ab} $ and divide the whole equation by $8\pi L_P^2$, to get:
\begin{equation}
 \int_\mathcal{R} \frac{d^3x}{8\pi L_P^2}  \, \sqrt{h}\, u_a g^{ij} \pounds_\xi N^a_{ij}
= \int_{\partial\mathcal{R}} \frac{d^2 x \, \sqrt{\sigma}}{L_P^2}\,\left( \frac{Na_\alpha r^\alpha}{4\pi}\right) - \int_\mathcal{R} d^3 x \, N \sqrt{h}\,( 2u^au^b \mathscr{F}_{ab})
\label{gravdyn}
\end{equation} 
As we said before, even in the general, time-dependent case, $r_i$ is in the direction of $a_i$ and we can interpret $T_{loc}=Na/2\pi$ as the Tolman redshifted Davies-Unruh temperature. So
the first term on the right hand side can be interpreted as an integral over $(dA/L_P^2)(1/2)k_BT_{loc}$.
In the second term, we identify $2N\mathscr{F}_{ab} u^a u^b = (\rho + 3p)N$ as the Komar energy density. (Note the factor 2  in the left hand side of this standard definition of the Komar energy density $(\rho+3p)$. Algebraically, this is exactly the factor 2 which makes everything consistent when we proceed from the heat content to the equipartition energy.)
Thus the  result in \eq{gravdyn} can be summarized in the form:
\begin{equation}
 \frac{1}{8\pi L_P^2} \int_\mathcal{R} d^3x \sqrt{h}\, u_a g^{ij} \pounds_\xi N^a_{ij} =  \epsilon \int_{\partial\mathcal{R}}\frac{d^2 x \, \sqrt{\sigma}}{L_P^2} 
\left( \half k_B T_{\rm loc}\right) - \int_\mathcal{R}d^3x\, \sqrt{h}\, \rho_{\rm Komar} 
\label{holoevl}
\end{equation} 

This result, again, has a remarkable physical meaning. If the spacetime is static and we choose the foliation such that $\xi^a$ is  the  Killing vector, then $\pounds_\xi N^a_{ij} =0$ and the left hand side vanishes. The equality of two terms on the right hand side can be thought of as representing the \textit{holographic equipartition} \cite{cqg04,a20,a21} if we define the bulk and surface degrees of freedom along the following lines: We take the number of surface degrees of freedom to be:
\begin{equation}
 N_{\rm sur}\equiv\frac{A}{L_P^2}=\int_{\partial \mathcal{R}} \frac{\sqrt{\sigma}\, d^2 x}{L_P^2}
\end{equation} 
 We next define an \textit{average} temperature $T_{\rm avg}$ of the boundary surface $\partial\mathcal{R}$ by:
 \begin{equation}
 T_{\rm avg}\equiv\frac{1}{A}\int_{\partial \mathcal{R}} \sqrt{\sigma}\, d^2 x\ T_{\rm loc}
\label{tav}
\end{equation} 
Next, we will define the bulk degrees of freedom $N_{\rm bulk}$ by the following procedure: \textit{If} the energy $E$ in the region $\mathcal{R}$ has reached equipartition at the average surface temperature $T_{\rm avg}$, \textit{then} $|E| = (1/2) N_{\rm bulk} k_B T_{\rm avg}$; that is, we define
 the number of bulk degrees of freedom by
\begin{equation}
N_{\rm bulk}\equiv \frac{|E|}{(1/2)k_BT_{\rm avg}}= \frac{\epsilon}{(1/2)k_BT_{\rm avg}}\int_\mathcal{R} \sqrt{h} d^3x\; \rho_{\rm Komar}
\label{nbulkgen}
\end{equation} 
where $E$ is the total Komar energy in the bulk region $\mathcal{R}$ contributing to gravity.
(The $\epsilon =\pm 1$  ensures that $N_{\rm bulk}$ remains positive even when the  Komar energy becomes negative.)  
This is the relevant value of $N_{\rm bulk}$ if we assume equipartition holds for the  
 energy $E$ in the bulk region  with the average surface temperature. 
Our result in \eq{holoevl} then says that \textit{comoving observers in any static spacetimes} will indeed find:
\begin{equation}
 N_{\rm sur} = N_{\rm bulk} \qquad ({\rm Holographic \ equipartion})
 \label{key1}
\end{equation}
That is, the equipartition is holographic  in all static spacetimes.
I have already discussed in Sec.~\ref{sec:1} how this result leads to $E=2TS$ for a Schwarzschild black hole where we have $N_{\rm sur}=(A/L_P^2)=4S; N_{\rm bulk}=E/((1/2)T)$ so that $N_{\rm sur} = N_{\rm bulk} $ is the same as $E=2TS$.

What is more, \eq{holoevl} suggests that \textit{the discrepancy from holographic equipartition --- resulting in a non-zero value for the right hand side --- is what drives the dynamical evolution of the spacetime.} We can write \eq{holoevl} as:
\begin{equation}
\int \frac{d^3x}{8\pi L_P^2}\sqrt{h} u_a g^{ij} \pounds_\xi N^a_{ij} = \frac{\epsilon}{2} k_B T_{\rm avg} ( N_{\rm sur} - N_{\rm bulk})
\label{evlnsnb}
\end{equation} 
Note that, even in a static spacetime non-static observers  will perceive a departure from holographic equipartition because \eq{holoevl}  --- while being generally covariant --- is  foliation dependent through the normal $u_i$. This is very clear if one considers the de Sitter spacetime which can be expressed as a static spacetime in  the Schwarzschild-like coordinates or as a time dependent spacetime in the Friedmann coordinates. Observers moving normal to  the static equal time surfaces will see that the de Sitter spacetime maintains holographic equipartition with $N_{\rm sur}= N_{\rm bulk}$. The observers moving normal to the equal time surfaces in the Friedmann coordinates, on the other hand, are geodesic observers with $a_i=0$. They will see that $N_{\rm sur}\neq  N_{\rm bulk}$ 
and in fact will find  that this discrepancy leads to the time dependence of the metric functions through \eq{evlnsnb}. Similar phenomenon can occur in any spacetime if we compare synchronous coordinates with other coordinates. This should cause no more surprise than the fact that, de Sitter spacetime, say, can be viewed as static or time dependent in two different foliations.

There are two other ways of rewriting $u_a g^{ij} \pounds_\xi N^a_{ij}$ in the left hand side of \eq{evlnsnb} by  relating it to   more familiar constructs in the Hamiltonian formulation of relativity \cite{adm}.
A straightforward computation  shows that the gravitational momentum flux $\sqrt{h} u_a g^{ij} \pounds_\xi N^a_{ij}$ has two other alternative
descriptions which are physically illuminating. We can show [see Appendix \ref{appen9.3}, \eq{ksquare} and \eq{niskpapp}] that:
\begin{equation}
 \sqrt{h} u_a g^{ij} \pounds_\xi N^a_{ij}=-h_{ab}\pounds_\xi p^{ab}; 
\quad p^{ab}\equiv\sqrt{h}(Kh^{ab}-K^{ab})
\label{niskp}
\end{equation} 
and
\begin{equation}
 \sqrt{h} u_a g^{ij} \pounds_\xi N^a_{ij}=2N\sqrt{h} [K_{ij} K^{ij}  -u^a \nabla_a K]
\end{equation} 
So we can rewrite \eq{evlnsnb} either as
\begin{equation}
 \int \frac{d^3x}{4\pi L_P^2}N\sqrt{h} [K_{ij} K^{ij}  -u^a \nabla_a K] = \frac{\epsilon}{2} k_B T_{\rm avg} ( N_{\rm sur} - N_{\rm bulk})
\end{equation} 
or, more simply, as:
\begin{equation}
 -\int \frac{d^3x}{8\pi L_P^2}h_{ab}\pounds_\xi p^{ab}  = \frac{\epsilon}{2} k_B T_{\rm avg} ( N_{\rm sur} - N_{\rm bulk})
\end{equation} 
which clearly shows how the departure from holographic equipartition drives the evolution of geometry \cite{tpraa}, providing an interesting alternative description of spacetime dynamics. 

These results provide a direct and simple relation between the dynamical evolution of the spacetime geometry and the departure from thermodynamic equipartition. It also encodes the holographic nature of gravity (noticed earlier in several contexts, like e.g., in  \cite{a22,ayan,a24}) by introducing the surface and bulk degrees of freedom of a region and their equality in static spacetimes. On the left hand side, we essentially have a measure of  the change in the  gravitational momentum.  Given the correspondence between $\delta N^a_{bc}$ and $\delta T$, one should be able to interpret the left hand side as the `heating of  the spacetime' \cite{a25,a26}.  One also notices the obvious connection with the membrane paradigm \cite{a27,a28,a29}  which brings in a viscous tensor and the concept of ``dissipation without dissipation'' (introduced in e.g., Sec. 2.3 of \cite{dis},\cite{a25}).  Further work is required to make these connections precise.

It is also gratifying that the result involves the rate of change of $p^{ab}$ which encodes the dynamical content of the field equations in the following sense: In the standard Hamiltonian formulation of general relativity, one has two constraint equations $\mathcal{H}=0,\mathcal{H}_\alpha=0$, one equation for $\pounds h_{ab}$ which defines $p_{ab}$ in terms of $h_{ab}$, and one dynamical equation involving $\pounds p_{ab}$. What we get here is the trace part of  this dynamical equation.
This is understandable, because
to derive the  result in \eq{evlnsnb} I only used the constraint equation $R_{ab}u^au^b=(8\pi L_P^2)\mathscr{F}_{ab}u^au^b$ and we cannot get out more information than we put in. But, as is well-known (see e.g., p. 259 of \cite{tpgravity}), the validity of the constraint equation for all observers leads to the full set of field equations of gravity. In this sense, we only need the validity of \eq{evlnsnb} for all regions of space and all foliations  in order to obtain the full set of field equations. It should be possible to ``reverse-engineer'' the above result and do this, but I believe that  it is not going to add more insights into the situation. (I will describe in Sec.~\ref{sec:6} how to obtain the gravitational field equations from a suitable thermodynamic variational principle properly.)

As a description of  the gravitational dynamics, we are essentially comparing \eq{evlnsnb} with the standard equation $R_{ab}=(8\pi L_P^2)\mathscr{F}_{ab}$. My main problem with the latter is that it is very difficult to attribute \textit{physical} meaning to the left hand side. Further, there is no natural separation of time-dependent and static geometries and it is not  obvious what drives the spacetime evolution. It is \textit{not} $T_{ab}$ because: (a) One can have time dependent solutions when $T_{ab}=0$ and static solutions when $T_{ab}\neq 0$. (b) The notion of time independence, while can be expressed quite geometrically in terms of existence of a timelike Killing vector, is not easy to translate in terms of actual metric functions. As I have said several times, the natural coordinate system used by two different sets of observers can have a time independent metric in one case and a time-dependent metric in the other.

The description in \eq{evlnsnb} does better as regards these issues. 
First, it brings in the observers through the foliation and --- contrary to what is sometimes thought --- \textit{this is a very desirable feature}. \textit{All} thermodynamics \textit{is} observer dependent (see Sec. 4 of\cite{a9}) and any attempt to re-interpret the field equations in the thermodynamic language has to acknowledge this feature up front. [Of course, once we demand the validity of \eq{evlnsnb} for all foliations, we will get back $R_{ab}=(8\pi L_P^2)\mathscr{F}_{ab}$ which has no trace of foliation left in it. But the basic equation, viz. \eq{evlnsnb} has a foliation dependence.] 

Second, for a given observer, 
we now have a natural separation of static and dynamic spacetimes in terms of holographic equipartition. In stationary geometries, we can choose $\xi^a$ to be the timelike Killing vector and we immediately get $N_{\rm sur}=N_{\rm bulk}$. When this condition does not hold, we get time evolution due to departure from holographic equipartition. 
This result also provides a physically transparent statement about dynamics. One should also mention in this context that the holographic thermodynamics necessarily involves a nonlocal description because of the involvement of a surface and bulk.

When there is no matter, our \eq{holoevl} reduces to an interesting form:
\begin{equation}
 \frac{1}{8\pi L_P^2} \int d^3x \sqrt{h}\, u_a g^{ij} \pounds_\xi N^a_{ij} =  \epsilon \int_{\partial\mathcal{R}}\frac{d^2 x \, \sqrt{\sigma}}{L_P^2} 
\left( \half k_B T_{\rm loc}\right) 
\label{tab0}
\end{equation} 
This equation holds, among other contexts, in the description of a gravitational wave spacetime. It shows that the momentum change in a region containing a source-free gravitational field is directly related to the surface heat energy. 
The comments made earlier regarding the membrane paradigm \cite{a27,a28,a29} and the  concepts like  `heating of spacetime' \cite{a25,a26},   ``dissipation without dissipation'' (introduced in Sec. 2.3 of \cite{dis};\cite{a25}) are again applicable in this context.
I hope to get back to the implications of this description in a future work.

\subsection{An explicit example}

As a concrete example of the above ideas, let us consider a time dependent, spherically symmetric, metric of the form
\begin{equation}
 ds^2 = - V dt^2 + W dr^2 + r^2 d\Omega^2
\end{equation} 
where $V$ and $W$ are functions of $r$ and $t$. In addition to standard black hole spacetimes and collapse scenarios, \textit{this form of the metric also describes all the Friedmann universes}, since all of them are spherically symmetric and (in general) time dependent. The purpose of this short section is to show how the balance between the three terms in \eq{holoevl} is achieved in such a concrete case. My discussion will be brief and applications to cosmology will be discussed elsewhere.

The acceleration computed using 
$a^i = ( \gu ij + u^i u^j) \nabla_j  \ln \sqrt{V}$
gives $a^r = (V'/2VW)$ with the magnitude $Na = (V' / 2\sqrt{VW})$ so that
\begin{equation}
 D_\alpha ( 2 N a^\alpha) = \frac{1}{r^2 \sqrt{W}}\ \partial_r \left( r^2 \frac{V'}{\sqrt{VW}}\right)
\label{dsix}
\end{equation} 
(We use a  prime to denote the  derivative with respect to $r$ and an overdot to denote the  derivative with respect to $t$.)  Another explicit computation gives
\begin{equation}
  2N R_{ab} u^a u^b = \frac{2N}{N^2} R_{00} = \frac{2}{\sqrt{V}} \left( \frac{V''}{2W} - \frac{V'W'}{4 W^2}+\frac{V'}{Wr} - \frac{(V')^2}{4VW} - \frac{\overset{\bm{\centerdot\centerdot}}{W}}{2W} + \frac{\overset{\bm \centerdot}{\,W^2}}{4 W^2} + \frac{\overset{\bm \centerdot}{V} \overset{\bm \centerdot}{W}}{4 V W}\right)
 \label{dseven}
\end{equation} 
We now see that the first four terms within the bracket in \eq{dseven} will survive even in the static limit, while the remaining four terms have time derivatives. On the other hand, the expression in \eq{dsix} has no time derivatives and we know that \eq{dsix} and \eq{dseven} \textit{have to match} in the static limit. This is indeed the case because, on expanding the right hand side of \eq{dsix}, one can easily show that 
\begin{equation}
 D_\alpha ( 2 N a^\alpha) = \frac{2}{\sqrt{V}} \left\{\frac{1}{2} \frac{V''}{W}  - \frac{1}{4} \frac{V'W'}{W^2} +
 \frac{V'}{rW}  - \frac{1}{4} \frac{(V')^2}{VW}\right\}
\end{equation} 
Clearly  these terms --- which remain the same even in the time dependent case --- match with the first four terms in \eq{dseven}, allowing us to write
\begin{equation}
 2N R_{ab} u^a u^b = D_\alpha ( 2 N a^\alpha) + \frac{2}{\sqrt{V}} \left\{ \frac{\overset{\bm \centerdot}{V}\overset{\bm \centerdot}{W}}{4 VW} + \frac{\overset{\bm \centerdot}{\,W^2}}{4 W^2} -  \frac{\overset{\bm{\centerdot\centerdot}}{W}}{2W} \right\}
\end{equation} 
We therefore identify, from our general result, that
\begin{equation}
 - u_a \gu ij \pounds_\xi N^a_{ij} = \frac{2}{\sqrt{V}} \left\{\frac{\overset{\bm \centerdot}{W}}{4 W}\frac{(VW)^{\overset{\bm \centerdot}{^{}}} }{(VM)} -\frac{\overset{\bm{\centerdot\centerdot}}{W}}{2W}  \right\}
=-\frac{1}{\sqrt{W}}\frac{\partial\ }{\partial t} 
\left(  \frac{\overset{\bm \centerdot}{W}}{\sqrt{VW}}
                     \right)
\end{equation} 
which is an explicit expression for $u_a \gu ij \pounds_\xi N^a_{ij}$ in this simple context.
For the sake of completeness, one can verify the last expression explicitly. In this particular case, we have
$K_{\mu\nu} = - (1/2N) \partial_t h_{\mu\nu}$, leading to $K_{rr} = - (1/2\sqrt{V})\dot W$ and $K=- (1/2\sqrt{V})(\dot W/W)$. 
Working out the expression for $2N (K_{\mu\nu} K^{\mu\nu} - u^a \partial_a K)$, one can explicitly verify that 
it matches with $u_a g^{ij} \pounds_\xi N^a_{ij}$.
Thus, the static terms of $2N R_{ab} u^a u^b$ match with $D_\alpha (2N a^\alpha)$ while the time dependent terms match with 
$ - u_a g^{ij} \pounds_\xi N^a_{ij}$.

\section{Gravitational Energy and the surface thermodynamics}\label{sec:5}

I will now demonstrate yet another curious connection between bulk gravitational dynamics and surface thermodynamics by showing that, when the field equations hold, there is a natural notion of total energy of matter plus gravity which is equal to the surface heat content. To do this, we  first have to define a gravitational four momentum flux $P^a$ such that its integral over $u_a\sqrt{h}d^3x$ gives a sensible notion of gravitational energy. The usual folklore is that such notions are either impossible to define or (when defined) fairly useless. We will see that the thermodynamic considerations change the picture quite a bit.

\subsection{Bulk total energy matches the surface heat energy}

It is, of course, well-known that one cannot define a covariant, local, notion of gravitational energy density. Further, the very notion of energy or energy density, operationally speaking, has an implicit observer dependence. For example, a particle can have a covariant, observer independent, four-momentum $\mathbf{p}$; but the energy attributed to this particle by an observer will be $E=-\mathbf{p}\cdot\mathbf{u}$ where $\mathbf{u}$ is the four-velocity of the observer. Taking a cue from this we will use an \textit{extra vector field $q^a$} to define the gravitational four momentum density $P^a_H[q]$ \textit{associated} with it. (The subscript $H$ is to remind us that we are using the Einstein-Hilbert Lagrangian; later on, in Sec.~\ref{sec:7.2}, we will discuss what happens when we use other forms of  the Lagrangian. When no subscript is used, it means the  Einstein-Hilbert Lagrangian.) Of course, such an entity depends on an extra vector field and not an intrinsic property of  the spacetime geometry alone; however, we will see that it has a natural interpretation when the vector is chosen to be the time-development vector $\xi^a$. In the past such ideas have been explored with $\zeta^a$ (for a small sample see e.g., \cite{defenergy,qle} and references therein) but, again, we will see that the thermodynamic interpretation works better with $\xi^a$.

The exact form for $P^a_H[q]$ 
 associated with a vector field $q^a$ has already been motivated in Sec.~\ref{sec:2.1} and we have defined the vector in \eq{defpa1} as
\begin{equation}
 P^a(q) \equiv g^{ij} \pounds_q N^a_{ij} + L q^a
\end{equation} 
The Noether current approach  allows us to study this $P^a$ quite easily. To begin with, we have the relation:
\begin{equation}
 \nabla_b J^{ab} = 2 R^a_b q^b + g^{ij} \pounds_q N^a_{ij} = 2 G^a_b q^b + Lq^a + g^{ij} \pounds_q N^a_{ij} \equiv 2 G^a_b q^b + P^a
\end{equation} 
from which we get an alternative, and more useful, definition:
\begin{equation}
 P^a(q) = g^{ij} \pounds_q N^a_{ij} + L q^a=\nabla_b J^{ab}-2 G^a_b q^b
\label{defpa}
\end{equation} 
It is also easy to see that 
\begin{equation}
 \nabla_a P^a = - 2 \nabla_a(G^a_b q^b) = - 2 G^{ab}\nabla_a q_b 
\end{equation} 
which vanishes on-shell in pure gravity making $P^a$ a conserved current in this context. In fact, $P^a$  is what is usually called Noether current in the literature, obtained from our Noether current $J^a$ by dropping terms that vanish on-shell. However, the idea that  the Noether current is actually a measure of gravitational momentum flux does not seem to have been emphasized (and explored) in detail. We have seen enough evidence already to realize that $J^a$ is directly related to energy (rather than, for example, entropy; see e.g.,\cite{a40,jise}) and results in this section will reinforce this idea. 

All this is true for general $q^a$. Let us now specialize to $q^a=\xi^a=Nu^a$ when we will be able to obtain several interesting results.
Taking the dot product of $P^a(\xi)$ with $u_a$ where $u_a = - N \delta_a^0$ and using \eq{fin1}, we get:
\begin{equation}
 u_a P^a(\xi) = - 2NG_{ab} u^au^b + u_a J^a[\xi] = -2 NG_{ab} u^a u^b + 2D_\alpha (Na^\alpha)
\end{equation} 
On the other hand, the corresponding matter energy density is given by $NT_{ab}u^au^b$. Adding it to gravitational part, using the constraint equation $2G_{ab}u^au^b=T_{ab}u^au^b$, and integrating the expression over a three dimension region bounded by an $N=$ constant surface (as we did in the previous sections), we get a  simple expression for the  total (matter+gravity) energy in a bulk region on shell:
\begin{equation}
 \int_\mathcal{R} d^3x\sqrt{h}u_a[P^a(\xi)+NT^a_bu^b]=\int_{\partial\mathcal{R}} d^2x Ts
\label{totalbulk}
\end{equation} 
That is, when the gravitational field equations are satisfied, the total energy in the bulk region is exactly equal to the heat content of the bounding surface. This is yet another connection between gravitational dynamics, holography and surface thermodynamics. We can also say that this total energy is \textit{half} of the thermal, equipartition, energy of the boundary obtained by attributing $(1/2)k_BT$ energy to each of the $(A/L_P^2)$ degrees of freedom.

The expression for gravitational energy density itself has some interesting features. Note that the quantity 
\begin{equation}
 2 G_{ab} u^a u^b = K^2 - K_{ab} K^{ab} + {}^3R = - \mathcal{H}_{\rm adm}
\end{equation} 
is essentially the 00 bulk part of the  ADM Hamiltonian defined by the second equality above
(see e.g., Sec.  12.4.1 of \cite{tpgravity}).
It therefore follows that $u_a P^a[\xi] $ is essentially $\mathcal{H}_{\rm adm}$ plus an \textit{important} total divergence:
\begin{equation}
 u_a P^a[\xi] = N\mathcal{H}_{\rm adm} + D_\alpha (2Na^\alpha) 
\label{uph}
\end{equation}
The integral of this expression over a three surface with measure $\sqrt{h}d^3x$ will lead to the surface heat content from the second term and the bulk energy from the first term:
\begin{equation}
 \int_\mathcal{R} d^3x\sqrt{h}u_aP^a(\xi)=\int_\mathcal{R} d^3x\sqrt{h}N\mathcal{H}_{\rm adm} +\int_{\partial\mathcal{R}} d^2x Ts
\end{equation} 
When the constraint equation holds, the  sum of $\mathcal{H}_{\rm adm}$ and the matter energy density $H$  vanishes, which leads to the surface heat content as the final result.

\subsection{Variation of the gravitational energy}

Since the gravitational energy defined through our $P^a$ seems to lead to interesting results, it is useful to study its properties and ---  in particular --- investigate how it changes due to processes operating on the boundary. In fact, the 
current $P^a(q)$ (for an arbitrary diffeomorphism with vector field $q^a$) also allows us to introduce a symplectic structure for gravity which is well-known in the  literature (for a small sample, see e.g. \cite{symp, bp1}). 
I will show that, these results also have an interesting thermodynamic interpretation.

The symplectic structure can be obtained by varying $\g P^a$ and manipulating the resulting terms. We then find that, for arbitrary variations with fixed $q^a$, we have the result [see Appendix \ref{appen9.4}, \eq{varPapp}]:
\begin{equation}
 \delta ( \g P^a_H) - q^a R_{ij} \delta f^{ij}=\g \omega^a +\partial_c( f^{lm}\delta N^{[a}_{lm} q^{c]})
 \label{varP}
\end{equation}
where we have defined the symplectic form as
\begin{equation}
 \g \omega^a (\delta,\pounds_q)\equiv \delta f^{lm} \pounds_q N^a_{lm} - (\pounds_q f^{lm}) \delta N^a_{lm}
\end{equation}
This expression is completely general and $\omega^a$ involves one arbitrary variation $\delta$ and one Lie derivative $\pounds_q$.

We can use this result to obtain the change in the total gravitational energy as the system evolves, in terms of a Lie derivative along $\xi^a$.
To do this, we  need to compute $\pounds_\xi(\sqrt{h}u_aP^a_H(\xi))$ which is straightforward. First, we obtain from \eq{varP} an expression for the variation of $\delta(\sqrt{h}u_aP^a_H(\xi))$  (after setting $q^a=\xi^a$), either by using the fact that $u_a=-N\nabla_at$ and taking the dot product or working it out from the basic definition. The result is [see Appendix \ref{appen9.4}, \eq{varPuapp}]:
\begin{equation}
 \delta \left( \sqrt{h}\, u_a P^a_H(\xi)\right) + R_{ab} \delta f^{ab} = \sqrt{h}\, \omega^a (\delta, \pounds_\xi) u_a + \partial_c \left[ h^c_a f^{lm} \delta N^a_{lm}\right]
\end{equation} 
which is  valid for arbitrary variations.
Next we evaluate the above expression when the variation is due to a diffeomorphism along $\xi^a$ so that we can replace $\delta$ by $-\pounds_\xi$. Because of the antisymmetry, the $\omega^a$ term drops out and we obtain [see Appendix \ref{appen9.4}, \eq{varPuapp1}]:
\begin{equation}
 \pounds_\xi \left( \sqrt{h}\, u_a P^a_H(\xi)\right)   = \partial_c \left[ \left( 2 G^c_b \xi^b + h^c_a \gu lm \pounds_\xi N^a_{lm}\right) \g \right]
\end{equation} 
Integrating over a region of space, we get:
\begin{equation}
  \pounds_\xi H_{\rm grav} \equiv \pounds_\xi \int_\mathcal{R} d^3 x\, \sqrt{h}\, u_a P^a = \int_{\partial\mathcal{R}} d^2 x \, \sqrt{\sigma}\, N r_a \left( T^{ab} \xi_b + \gu lm \pounds_\xi N^a_{lm}\right)
\label{energychange}
\end{equation} 
where we have also used the field equations $2G_{ab}=T_{ab}$.
This expression shows how the energy in the bulk changes due to processes taking place at the boundary.
(The result in \eq{varP} was obtained under the assumption $\delta q^a=0$ but it will continue to hold if the variation $\delta$ is due to a diffeomorphism along the same vector field $q^a$ because $\pounds_qq^a=0$  will make all the additional terms vanish. This can be verified explicitly.)
Since we have already established that the energy is given by a boundary integral in \eq{totalbulk}, it is obvious that we will only get a surface contribution for $\pounds_\xi H_{\rm grav}$. Of the two terms in \eq{energychange}, one is clearly due to the  matter energy flux across the boundary. The second term, which is more interesting to us, is again the ubiquitous $f\delta N$ term. We once again see that this can be thought of as the change in the gravitational heat due to the processes at the boundary. 
In fact, if we consider pure gravity, this result becomes:
\begin{equation}
  \pounds_\xi H_{\rm grav}  = \int_{\partial\mathcal{R}} d^2 x \, \sqrt{\sigma}\, N r_a \left( \gu lm \pounds_\xi N^a_{lm}\right)
\label{hdot}
\end{equation}
which is directly applicable to, say,  gravitational waves. We find that the energy associated with a gravitational wave metric in a bulk region changes due to surface processes involving $\pounds_\xi N^a_{lm}$. This is similar to the result we obtained earlier for the dynamical evolution of pure gravity in \eq{tab0} and should have an interpretation as the heating of  the surface \cite{a25,a26} in terms of  the surface viscous tensor that arises in the membrane paradigm \cite{a27,a28,a29} and the ``dissipation without dissipation'' (Ref. \cite{a25}, sec.2.3 of Ref. \cite{dis}). I have already commented about this connection earlier and the mathematical results in \eq{tab0} and \eq{hdot} essentially describe the same physics. 

We will conclude this section after commenting on the relation of these results to some other well-known results in the literature.

The symplectic form and related constructs also arise in the approach to compute horizon entropy using the central charge and Virasoro algebra, pioneered by Carlip \cite{carlip,bp1}. This approach requires us to start with $P^a_H(q)$ defined for a vector $q^a$, and construct its Lie derivative along another vector $v^a$ to obtain $\pounds_vP^a[q]$. This allows a natural definition of the symmetric ($ \pounds_vP^a[q]+\pounds_qP^a[v]$)   and the antisymmetric ($ \pounds_vP^a[q]-\pounds_qP^a[v]$) combinations of the double variations. From the above analysis, we see that one can do the same with the energy density expression. The symplectic form vanishes in the symmetric expression, leading essentially to the surface term. It is this surface term which contributes to the central charge construction and to the horizon entropy \cite{carlip,bp1}. It has been shown in a previous work \cite{bp2} that one can obtain the same result by working essentially with the surface term in the action and performing the necessary algebra. The  results obtained in this paper show that these are fairly generic features  arising from the fundamental relation between the total energy and the boundary heat energy.

We also mention briefly the connection between these boundary terms and the ones usually encountered in the conventional Hamiltonian formulation of general relativity. 
Using the second relation in \eq{defpa}, we can also relate this variation of $P^a_H$ to the variation of the Noether potential. A simple calculation gives [see Appendix \ref{appen9.4}, \eq{deljab1}],
when the original spacetime is on-shell with $G_{ab}=0$ but $\delta G_{ab}\neq0$, the result:
\begin{equation}
  \partial_b \left\{ \delta (\g J^{ab}_H) -  \g g^{lm}\delta N^{[a}_{lm} q^{b]}\right\} = \g\omega^a + 2 \delta (G^a_b q^b \g ) 
  \label{deljab}
\end{equation} 
This allows us to make a connection with the usual Hamiltonian formulation of the theory.
In the usual approach, one deals (see e.g., \cite{hvar})
with the integral of the variation of $\delta(\mathcal{H} N+ \mathcal{H}^\alpha N_\alpha)$ (where $\mathcal{H}$ and $\mathcal{H}^\alpha$ are the constraints essentially corresponding to $G^0_0$ and $G^0_\alpha$) which will lead to a bulk term \textit{and a surface contribution} in the form:
\begin{equation}
\delta \int_\mathcal{R} d^3 x \left( N\mathcal{H} + N_\alpha \mathcal{H}^\alpha\right) = \int_\mathcal{R} d^3x\, \delta B + \int_ {\partial\mathcal{R}} d^2x\, \delta S 
\label{eq4}
\end{equation} 
The variation in the left hand side is  the same as $-\delta(2G_{ab}u^aq^b)$ if we choose $q^a=\zeta^a=Nu^a+N^a$. Further, the bulk term in the above variation, on the right hand side, has to cancel the bulk term involving the integral of $u_a\omega^a$ when the equations of motion hold. (In fact, this is how one obtains the equations of motion in the symplectic approach.) The consistency of \eq{deljab} then requires that the surface term in \eq{deljab} \textit{should} match with the surface term in \eq{eq4}. That is, we have the result:
\begin{equation}
 \int_{\partial\mathcal{R}} d^2x\, \delta S = \int_{\partial\mathcal{R}} d^2 x\, r_b u_a \left( \delta (\g J^{ab}) - f^{lm} N^{[a}_{lm} \zeta^{b]} \right)
\end{equation} 
This allows a simple way to identify and interpret the surface term in the variation of the Hamiltonian. While this result is known previously in the literature, our approach suggests that even the conventional Hamiltonian formulation of gravity can be possibly recast in a thermodynamic language.

\section{Where did we go wrong with gravity?}\label{sec:6}

The results described above, as well as several other pieces of work  in this area, suggest an intriguing connection between: (i) the thermodynamics attributed to null surfaces by local Rindler observers and (ii) dynamics of gravity. But in the description given above, I have only \textit{re-interpreted} the standard gravitational dynamics --- \textit{which, I think, is actually flawed} --- in the thermodynamic language. I will now show how it is possible to provide a completely independent, stand-alone,  derivation of the \textit{correct} gravitational field equations from a thermodynamic perspective.

\subsection{The single most important fact about the gravitational dynamics}\label{sec:6.1}

To motivate this, I begin by stressing the \textit{single most important fact} about gravitational dynamics which --- purely because of historical accident --- is  completely ignored in formulating gravitational field equations: \textit{Gravity does not couple to  the bulk energy density arising from the addition of a constant to the  matter Lagrangian}. Any attempt to describe gravity without incorporating this observed feature is bound to be wrong.

This fact, in turn, requires that the  gravitational field equations \textit{must be} invariant under the symmetry transformation of the matter sector equations:
\begin{equation}
L_{\rm matter} \to L_{\rm matter}+ \mathrm{constant},
\label{lmsym}
\end{equation}  
resulting in $T^a_b \to T^a_b + ({\rm constant})\ \delta^a_b$. (The electroweak symmetry breaking, for example, is equivalent to  the shifting of the standard model Lagrangian by a large constant and we know that the evolution of the universe was unaffected by this transition.) 
The standard gravitational field equations, in contrast to matter field equations, are \textit{not} invariant under the addition of a constant to the matter Lagrangian.
The addition of the constant changes the energy-momentum tensor of the matter by 
\begin{equation}
 T^a_b \to T^a_b + ({\rm constant}) \ \delta^a_b
 \label{tsym}
\end{equation} 
Clearly, the gravitational field equations now become $\mathcal{G}^a_b = T^a_b + ({\rm constant}) \ \delta^a_b$ which is equivalent to the introduction of a \cc\, if one was not present originally, or changing its numerical value, if a \cc\ was originally present in the gravitational Lagrangian. Obviously, this  is the crucial problem related to the \cc, viz., that its numerical value (either zero or non-zero) can be altered by the transformation in \eq{lmsym} which leaves the matter equations unchanged. A particle physicist interested in the standard model can choose the overall constant in the matter Lagrangian arbitrarily because the standard model does not care for this constant. But each choice for this constant will lead to a different value for the cosmological constant and a different geometry for the universe, many of which, of course, are observationally untenable.

Another way of stating this problem is as follows: Suppose we discover a fundamental principle which allows us to determine the numerical value of the cosmological constant (either zero or non-zero). Such a principle cannot help us if the gravitational field equations are not invariant under the transformations in \eq{lmsym} or \eq{tsym}.  
The above discussion allows us to  identify three  ingredients which are \textit{necessary} to solve the \cc\ problem: 
\begin{enumerate}
 \item The gravitational field equations \textit{must be} made invariant under the transformations in \eq{lmsym} and \eq{tsym} so that gravity is ``protected'' from the shift in the zero level of the energy densities.
 
 \item At the same time, the solutions to the field equations \textit{must} allow the cosmological constant to influence the geometry of the universe, because without it we cannot possibly explain the observed accelerated expansion of the universe. 
 
 \item We need a fundamental physical principle to determine the numerical value of the \cc\ since it cannot be introduced as a low energy parameter in the Lagrangian if the theory is invariant under the transformation in \eq{lmsym}.
\end{enumerate}

The first two requirements might at first sight sound impossible to satisfy simultaneously, but it can be achieved! The trick is to construct a set of gravitational field equations which are invariant under the transformation in  \eq{tsym}  but allow the inclusion of a \cc\
as an \textit{integration constant} in the solutions. As an example, consider a theory in which the field equations are given by the requirement:
\begin{equation}
 (\mathcal{G}^a_b - T^a_b) n_a n^b = 0
 \label{tf}
\end{equation} 
for all \textit{null} vectors $n^a$ in the spacetime \cite{aseemtp}. 
Here $\mathcal{G}^a_b = 2 G^a_b$ in Einstein's theory and could be some other geometrical tensor in alternate theories of gravity, but necessarily satisfying the generalized Bianchi identity $\nabla_a \mathcal{G}^a_b=0$.
The above equations can be solved  by $\mathcal{G}^a_b - T^a_b = F(x) \delta^a_b$, but the generalized  Bianchi identity
($\nabla_a \mathcal{G}^a_b =0$)
and the conservation of the energy-momentum tensor $(\nabla_a T^a_b =0)$ imply that $F(x)$ must be a constant. Therefore, \eq{tf} is equivalent to the standard gravitational field equations with an arbitrary \cc\ appearing as an integration constant. Thus, if we can construct a theory of gravity in which the field equations reduce to those in \eq{tf}, then we would have achieved the first two requirements in the list for solving the \cc\ problem. 

This turns out to be an extremely strong demand and has important consequences often overlooked in attempts to ``solve'' the \cc\
problem. To see this, consider any theory of gravity interacting with matter satisfying the following three conditions: 
\begin{enumerate}
 \item The theory is generally covariant so that the matter action is constructed by integrating  a scalar Lagrangian $L_m(g_{ab},\phi_A)$ over the measure $\sqrt{-g} d^4x$. 
 \item The matter equations of motion are invariant under the transformation $L\to L + C$ where $C$ is a scalar constant.
 \item The gravitational field equations are obtained by an \textit{unrestricted} variation of the metric tensor $g_{ab}$ in the total action obtained by integrating a local Lagrangian over the  spacetime.
\end{enumerate}
It is easy to see that we \textit{cannot} solve the \cc\ problem in any theory satisfying the above three requirements. (This was clearly emphasized in Section IV of ref. \cite{tpap}). In particular, one cannot obtain the gravitational field equations of the form in \eq{tf} in any theory which satisfies the above three criteria.

We thus see that even though all the three criteria stated above seem very reasonable, they together will prevent us from solving the \cc\ problem; so we need to give up at least one of them. Assuming we do not want to give up general covariance of the theory or the freedom to add a constant to the matter Lagrangian, we can only tinker with the third requirement. 

One simple way of obtaining \eq{tf} is to postulate that the gravitational field equations are obtained by varying the metric but keeping $\sqrt{-g} = $ constant. Such theories, called unimodular theories of gravity 
--- involving only a restricted variation of the metric to bypass the condition (3) above ---
 have been studied in the literature in the past \cite{unimod}. Unfortunately, the motivation to keep $\sqrt{-g} = $ constant is at best weak and at worst non-existent.  

It is also possible to obtain \eq{tf} from an alternative perspective of gravity which treats gravity as an emergent phenomenon (see ref.\cite{aseemtp}). In this approach, one associates thermodynamic potentials with all null vector fields in a spacetime. Maximization of the relevant thermodynamic potential (entropy, free energy, ...) associated with \textit{all} null vectors simultaneously will then lead to \eq{tf}. The maximization involves varying the null vector fields rather than the metric and hence it bypasses  the third requirement in our list.
The metric is not varied at all to obtain the field equations.  In such an approach,  the original variational principle itself (not just the field equations) is invariant under the transformation in \eq{tsym}.\footnote{For a complete solution to  the \cc\ problem, we need, in addition to this, a physical principle to determine its numerical value. Such a principle is described elsewhere \cite{hptp} and I will not discuss it here.} 

While this idea has been developed in Ref. \cite{aseemtp}, I will revisit it here for three reasons: (a) In the original work \cite{aseemtp},  we considered a functional obtained by integrating a scalar over \textit{spacetime} with the measure $\g d^4x$. It would be nicer to reformulate this idea using an integration over a \textit{null surface} rather than over spacetime. (b) The ideas developed in the previous sections of this paper suggest a connection between this formalism and the  Noether current, which I want to demonstrate. (c) The connection with the  Noether current, in turn, allows us to provide an interpretation of the functional by relating it again to a $f\delta N$ type structure which we have seen repeatedly. 

The first task of obtaining a variational principle based on a null surface is easy. A starting point for such a variational principle will be the following: Given  a null surface with an integration measure $d\lambda\,  d^2 x \sqrt{\sigma}$ and a null congruence $\ell^a$, let us construct the functional:
\begin{equation}
\mathcal{H}\equiv \int_{\lambda_1}^{\lambda_2} \frac{d\lambda\ d^2x}{16\pi}\,  \sqrt{\sigma}\, \left[-2R_{ab} +  16\pi T_{ab}\right] \ell^a\ell^b 
\end{equation} 
where we have reintroduced $16\pi G$ with $G=1$. 
It is then obvious  that 
extremising this functional with respect to \textit{all} $\ell^a$,  subject to the constraint $\ell^2=0$, 
 will lead to $R^a_b-8\pi T^a_b=f(x)\delta^a_b$ which, on using $\nabla_aG^a_b=0=\nabla_aT^a_b$ leads to Einstein's equations with an undetermined cosmological constant term arising as an integration constant. 
 
 A nicer version of the above variational principle can be obtained by noting that  for a null congruence $\ell^a$ on a null surface we have [see Appendix \ref{appen9.5}, \eq{1-1}]:
$R_{ab} \ell^a\ell^b  = - \nabla_i (\theta \ell^i) - \mathcal{S}$
where $\nabla_i \ell^i = \theta + \kappa$ and we have defined 
\begin{equation}
 \mathcal{S}\equiv [\nabla_i \ell^j \nabla_j\ell^i -(\nabla_i\ell^i)^2 ]
\end{equation} 
which we shall later identify with
 the heat (enthalpy) density associated with the null surface. 
 While integrating over a null surface
with the measure $d\lambda d^2 x \sqrt{\sigma}$,  we can ignore terms of the kind $\nabla_i(\phi \ell^i )$ (for any scalar $\phi$)  since they  produce only boundary contributions when $\ell^a$ is affinely parametrized. (See Appendix \ref{appen9.5}, \eq{bt1}).
It follows that we now have an alternative variational principle (in which we vary $\ell^a$) based on  the expression:
\begin{equation}
 Q\equiv \int_{\lambda_1}^{\lambda_2} \frac{d\lambda d^2x}{16\pi}\, \sqrt{\sigma}\, [2\mathcal{S} + 16\pi T_{ab}\ell^a\ell^b ]
\end{equation}
Since $T_{ab}\ell^a\ell^b$ can be thought of as the heat (enthalpy) density $(\rho+p)=Ts=TS/V$ of matter, we can again think of $(\mathcal{S}/8\pi)$ as essentially the heat density of the null surface. 

We will now show that these results can again be expressed in terms of appropriate Noether charges which lead to the heat density of the null surface in terms of $\pounds_\ell N^c_{ab}$. This provides the generalization of the earlier results to null surfaces.

\subsection{Heat density of the null surfaces}

Let $\ell_a$ be a null congruence defining a null surface which may not be affinely parametrized. 
If we take $\ell_a=A(x)\nabla_a B(x)$, then it is easy to prove that $\ell^i\nabla_i\ell_j=\kappa \ell_j$ where
$\kappa=\nabla_iA\nabla^iB=\ell^a\nabla_a\ln A$. Just as we computed the Noether current for $\xi^a$ while dealing with spacelike surfaces, we will compute the  Noether current for $\ell_a$ in the context of null surfaces. We get [see Appendix \ref{appen9.5}, \eq{4to2app} and \eq{null4to2}]:
\begin{equation}
 \ell_a J^a (\ell) = \nabla_b( \kappa\ell^b) - \kappa^2=\mathcal{D}_a (\kappa\ell^a) + \frac{d\kappa}{d\lambda}
 \label{4to2}
\end{equation} 
Note that the quantity $\nabla_b( \kappa\ell^b) - \kappa^2$ in the case of a null surface is analogous to the right hand side of $D_ia^i= D_\alpha a^\alpha =  \nabla_i a^i - a^2$ which arises when we deal with spacelike surfaces. Unfortunately the projection to a surface `orthogonal' to a null vector  is not well defined for us to introduce a covariant derivative. Instead, we have to work with a co-null vector $k_a$ defined such that $k^2=0$ and $k_a \ell^b =-1$. Then the projection to the 2-surface is provided by the projection vector $q^a_b = \delta^a_b + k^a\ell_b + k_b\ell^a$ and $\mathcal{D}_a$ is the covariant derivative defined using $q_{ab}$.
We thus get the final result for the Noether charge corresponding to the null congruence as
\begin{equation}
 \ell_a J^a (\ell) = 2 R_{ab}\ell^a\ell^b + \ell_a g^{ij} \pounds_\ell N^a_{ij} =  \mathcal{D}_a (\kappa\ell^a) +\frac{d\kappa}{d\lambda}
 \label{jforl1}
\end{equation} 
If we integrate this over the null surface with the measure $d\lambda d^2x\sqrt{\sigma}$ and ignore the pure boundary contribution, we get :
\begin{equation}
 \int d\Sigma_aJ^a (\ell)=\int d\lambda\ d^2x\sqrt{\sigma}\,\ell_a J^a (\ell) = \int d\lambda\ d^2x \sqrt{\sigma}\, \frac{d\kappa}{d\lambda}
\end{equation} 
This result is analogous to our result in \eq{flux3} and shows that --- in the case of null surfaces --- the Noether charge is again related to the `heating' of the boundary surface because $\kappa\propto T$.

The real importance of this result lies in the fact that it can be used in the variational principle introduced earlier based on the null surfaces. To do this,  we re-introduce the $16\pi G$ factor with $G=1$ and re-write \eq{jforl1} as
\begin{equation}
 - \frac{1}{8\pi} R_{ab}\ell^a\ell^b = \frac{1}{16\pi} \ell_a g^{ij} \pounds_\ell N^a_{ij} -  \frac{1}{16\pi}\left[  \mathcal{D}_a (\kappa\ell^a) +\frac{d\kappa}{d\lambda}\right]
\end{equation} 
 Adding the term $T_{ab}\ell^a\ell^b$ to both sides, integrating  over a null surface with the measure $d \lambda d^2x \sqrt{\sigma}$ and ignoring the surface contributions, we find that our variational principle can be based on the functional:
 \begin{equation}
\mathcal{Q} \equiv \int_{\lambda_1}^{\lambda_2} d\lambda\ d^2x\,  \sqrt{\sigma}\, \left[\frac{1}{16\pi} \left( g^{ij} \ell_a\pounds_\ell N^a_{ij}- \frac{d\kappa}{d\lambda}\right) +  T_{ab}\ell^a\ell^b \right]
\label{functional}
\end{equation} 
In other words, we can obtain the field equations with an arbitrary \cc\ by varying $\ell_a$ (with the constraint $\ell^2 =0$) in the above functional and assuming that variations  vanish at the boundaries ($\lambda = \lambda_1, \lambda_2$). 
Since $T_{ab}\ell^a\ell^b$ can be thought of as the heat (enthalpy) density $TS/V=Ts$ of matter, we can think of the rest as essentially the heat density of the null surface. 
The second term in the integrand of \eq{functional} can be written as 
\begin{equation}
 -\frac{1}{16\pi} \int d\lambda\ d^2x\,  \sqrt{\sigma}\, \frac{d\kappa}{d\lambda} 
= -\frac{1}{16\pi}  \int_{1}^{2} d^2x\,  \sqrt{\sigma}\, \kappa  + \frac{1}{16\pi}  \int d\lambda\ d^2x\, \kappa \frac{d \sqrt{\sigma}}{d\lambda}
\end{equation} 
where the first term contributes only at the two boundaries. Ignoring that, we find that the second term of \eq{functional}  is proportional to (in  units with $G=L_P^2$)
\begin{equation}
 \frac{1}{16\pi L_P^2} \int d^2x\,  \kappa  d \sqrt{\sigma} \propto  \int d^2x\,  T\, ds
\end{equation} 
where $T=\kappa/(2\pi)$ and the entropy density is $s = \sqrt{\sigma}/4$.  The structure of this term is  suggestive of the heating of the null surface.

When $\ell_a $ is affinely parametrized with $ \kappa=0$ (which is a choice we can always make), then the variational principle can  be based on the integral
\begin{equation}
 Q_1\equiv \int_{\lambda_1}^{\lambda_2} d\lambda\ d^2x\, \sqrt{\sigma}\, \left[\frac{1}{16\pi}g^{ij} \ell_a\pounds_\ell N^a_{ij} +  T_{ab}\ell^a\ell^b \right]
\end{equation}
This again shows that the quantity $g^{ij} \ell_a\pounds_\ell N^a_{ij}$ plays a vital role even in the derivation of  the field equations from an alternative extremum principle.
In fact, the field equations in the absence of matter can be obtained by extremising the expression
\begin{equation}
 Q_1\equiv \frac{1}{16\pi}\int_{\lambda_1}^{\lambda_2} d\lambda\ d^2x\, \sqrt{\sigma}\, \left[g^{ij} \ell_a\pounds_\ell N^a_{ij} \right]
\end{equation}
over all null surfaces simultaneously. We have seen earlier (\cite{KBP}; see 
\eq{stSdT})  that the integral of this term on a null surface has a very simple physical meaning in terms of the heat content of the null surface. So, at least for pure gravity, we can obtain a variational principle which has an interpretation in thermodynamic language.

\section{Variations on the basic theme}\label{sec:7}

So far we have worked with (i)  the  Einstein-Hilbert Lagrangian and  (ii) the vector field $\xi^a=Nu^a$ to obtain our results (except in the last section in which we used $\ell^a$). The ideas can be generalized in two obvious directions. 

First, one can consider Lagrangians related to  the Einstein-Hilbert Lagrangian by  total divergences. There are, again, two obvious choices: First is obtained by using the so called $\Gamma^2$ Lagrangian which is obtained by adding a total divergence to cancel  the surface term in \eq{bulksur} while the second is obtained by adding a $2K$ term at each boundary \cite{ghy}. This is equivalent  to adding a term  $\nabla_a(Ku^i)$ in the bulk (with similar terms for other timelike boundaries when required).

Second, one can consider vectors other than $\xi^a$ to study the properties of the Noether charge,  the Hamiltonian etc. with the obvious alternative being $\zeta^a=Nu^a+N^a$. The purpose of this section is to briefly describe what happens in these contexts.

\subsection{Noether currents for related Lagrangians}\label{sec:7.1}

We begin by considering the Noether current associated with $L_{\rm quad}$ which, as we know is not a scalar. But when two Lagrangians differ by a total divergence, their infinitesimal variations will also differ by a total divergence. Therefore, even if one of them is not a scalar, it will still be possible to obtain a conserved current associated with it. The strategy is to obtain the form of $\pounds_q (\g L_{\rm quad})$ in two different ways --- by explicit variation of $\Gamma$s and by using the functional variation with respect to the metric --- and equate the two expressions. The first calculation is considerably simplified by writing  $\g L_{\rm quad}=\g L_H-L_{\rm sur}$. The only care that is needed is in  the manipulation of the   Lie derivatives of non-tensorial objects [discussed in Appendix \ref{appen9.6}]. The explicit variation leads to the result [see Appendix \ref{appen9.6}, \eq{ib2app}]:
\begin{equation}
 \pounds_q(\g L_{\rm quad})
= \partial_a \left[ \g L_{\rm quad} q^a \right] -  \partial_a (\g K^a)
\end{equation} 
where
\begin{equation}
 K^a = g^{lm} \partial_l\partial_m q^a  - g^{al} \partial_l\partial_m q^m
\label{defK1}
\end{equation} 
On the other hand, performing the  variation of $\g\,  L_{\rm quad}$ by treating it as a functional of the metric, we have the result:
\begin{equation}
 \pounds_q(\g L_{\rm quad}) = \partial_a ( - 2 \g G^a_b q^b + N^a_{lm} \pounds_q f^{lm})
\end{equation} 
Equating the two, we get the conservation law $\partial_a (\g\, J^a_{\rm quad}) =0$ where
the Noether current for the quadratic Lagrangian may be defined as:
\begin{equation}
 \g J^a_{\rm quad} =  2 \g G^a_b q^b + \g L_{\rm quad}q^a - N^a_{lm}\pounds_q f^{lm} - \g K^a
\label{defjquad}
\end{equation} 
Using the standard result for the Noether current from Einstein-Hilbert action $J^a_H$ this can be expressed as [see Appendix \ref{appen9.6}, \eq{25novc5app}]:
\begin{equation}
J^a_{\rm quad} = J^a_H + \nabla_b ( V^{[a} q^{b]} ) 
= J^a_H + \nabla_b ( g^{lm} N^{[b}_{lm} q^{a]})
 \label{25novc5} 
\end{equation}
where we have used the result $ \g\, V^a = - f^{lm} N^a_{lm}$. If we write $J^a_H\equiv J^a_{\rm quad}+J^a_{\rm sur}$, thereby \textit{defining} a Noether current associated with $L_{\rm sur}$, then:
\begin{equation}
 J^{a}_{\rm sur} = - \nabla_b (V^{[a} q^{b]}) = \nabla_b (V^{[b} q^{a]})
=\nabla_bJ^{ab}_{\rm sur}; \quad J^{ab}_{\rm sur}\equiv V^{[b} q^{a]}=q^aV^b-q^bV^a
\end{equation} 

We see that  the  final result is remarkably simple \cite{bp1,bp2}: adding the divergence of an entity $v^a$ leads to an extra term in the Noether potential which is $(v^{[b} q^{a]})$. To understand this, at least when $v^a$ is a vector, let us consider the following question \cite{bp2}: Suppose we have a scalar Lagrangian $L=\nabla_aW^a$ which is total divergence of a vector $W^a$ so that the action is a pure surface term. What is the corresponding Noether current? We note that, treated as a scalar density,
\begin{equation}
\pounds_q(\g L)=\g \nabla_a(Lq^a)=\partial_a(\g Lq^a) 
\end{equation}
On the other hand, expanding out the Lie derivative, we have:
\begin{equation}
\pounds_q(\g L)=\pounds_q(\partial_a[\g W^a])=\partial_a[ \pounds_q(\g W^a)]
\end{equation}
where we have used the fact the Lie derivatives commute with partial derivatives since they are local variations. 
Equating the two, we get a conservation law  $\partial_a(\g J^a)=0$, where:
\begin{eqnarray}
 \g J^a&=&\g Lq^a-\pounds_q(\g W^a)\nonumber\\
&=&q^a\partial_b(\g W^b)
-[q^b\partial_b(\g W^a)-\g W^b\partial_bq^a+\g W^a\partial_bq^b]
\label{jsur1}
\end{eqnarray}
The first and  the third terms combine to give $\partial_b(\g q^a W^b)$ while the second and  the fourth combine to give $-\partial_b(\g q^b W^a)$. So we get \cite{bp2} the Noether current and potential  to be:
\begin{equation}
\g J^a= \partial_b(\g q^{[a} W^{b]})=\g \nabla_b(q^{[a} W^{b]});\quad
J^{ab}=q^aW^b-q^bW^a 
\end{equation}
So if $V^c$ was a vector we could have written down our  result in \eq{25novc5} using this fact; but in our case $V^c$ is not a vector. Nevertheless, we see that the non-vectorial part involving $K^a$ appears only \textit{additively} (because of the special form of $V^c$) and thus cancels in the evaluation of \eq{jsur1}.

The result obtained above allows us to tackle the second case, namely the one in which we have a $2K$ term on the boundary, with ease. 
Since this term arises from integrating $\nabla_a(Ku^a)$ in the bulk, it will lead to a Noether potential  $J^{ab}=K(q^au^b-q^bu^a)$ for the  diffeomorphism along a vector field $q^a$. (With more terms of similar nature included if there are other, say, timelike, boundaries). 

This fact --- in turn --- leads to a remarkable conclusion: \textit{The extra  term for the spacelike boundary vanishes for our special choice of vector $q^a=\xi^a$.} Therefore, we could have thought of  results in the previous sections, obtained for $J_H^a$, as holding also for  the $R+2K$ action. (This does not happen for $q^a=\zeta^a$.) 

The situation  regarding $L_{\rm quad}$ is different. Here, the Noether potential is $J^{ab}_{\rm sur}(u)=u^aV^b-u^bV^a$. While $2K=V^cu_c$ for a foliation with $N_\alpha=0$ (see p.250 of \cite{tpgravity}), in general $V^c$ can have a component orthogonal to $u^a$ which can contribute to $u^aV^b-u^bV^a$. This would bring in a nontrivial difference between $V^c$ and $K$ as far as Noether currents are concerned.
Also, note that  while the Noether potential for $K$ term is generally covariant, the one for $V^c$ is not.

\subsection{Gravitational Hamiltonian from $L_{\rm quad}$ and its variation}\label{sec:7.2}

The use of $L_{\rm quad}$ suggests an associated definition for gravitational momentum flux, which is sometimes used in the literature.
This expression, analogous to $p\dot q -L$ can be defined as 
\begin{equation}
P^a_{\rm quad} = N^a_{lm} \pounds_q f^{lm} - q^a \g L_{\rm quad}   
\label{gravmomflux}                                                             
\end{equation} 
It can be motivated exactly the way we did for $P^a_H$ (leading to \eq{defpa1}, but using $L_{\rm quad}$ instead of $L_H$. (The choice of overall sign is conventional and we could have switched it, as we did earlier). We will not repeat this argument.

The obvious trouble which makes the argument and motivation rather weak in this case, of course, is that $P^a_{\rm quad}$  --- in contrast to $P^a_H$ which we worked with all along --- is,  noncovariant due to the presence of $N^a_{bc}$ and $L_{\rm quad}$. In the literature, when it is used, one usually concentrates on its variation (which takes care of the noncovariance of $L_{\rm quad}$) and performs a background subtraction on $N^a_{bc}$ in its definition to make it covariant. These features make it somewhat less attractive than $P^a_H$, but we will briefly discuss it for the sake of completeness. 
 
 We can easily relate $P^a_H$ and $P^a_{\rm quad}$. 
Using the notation $\mathcal{V}^a=\g V^a$, one can show that [see Appendix \ref{appen9.7}, \eq{ib3app}]:
\begin{equation}
 \g P^a_H 
= - P^a_{\rm quad} - \left[ \g K^a + \partial_c ( q^{[c} \mathcal{V}^{a]})\right]
\end{equation} 
where $K^a$ is defined by \eq{defK1}.
This gives, on combining with the on-shell version of \eq{varP},   the  on-shell result:
\begin{eqnarray}
 \delta P^a_{\rm quad} &=& - \delta (\g P^a_H) - \delta (\g K^a) + \partial_c (q^{[c} \delta N^{a]}_{lm} f^{lm} + q^{[c} N^{a]}_{lm} \delta f^{lm})\nonumber\\
&=& -\g \omega^a - \delta (\g K^a) + \partial_c ( q^{[c} N^{a]}_{lm} \delta f^{lm})
\end{eqnarray}
The extra term $\delta(\g K^a)$ arises because of the non-tensorial character of $P^a_{\rm quad}$. As mentioned before, one way to avoid this is to redefine $P^a_{\rm quad}$ by changing $N^a_{lm} $ to $N^a_{lm} - [N^a_{lm}]_{\rm flat}$ in \eq{gravmomflux} and use the result in Appendix \ref{appen9.6}, [\eq{lie0}] to make everything covariant. This will result in $\delta(\g K^a)$ disappearing in the above expression at the price of requiring a background subtraction.

We can also obtain corresponding expressions for $\delta ( \g J^a_{\rm quad})$ by relating it to $\delta (\g J^a_H)$ using  \eq{25novc5}. Then we get [see Appendix \ref{appen9.7}, \eq{ib4app}]
\begin{equation}
 \delta ( \g J^a_{\rm quad}) 
= \g \omega^a + \partial_b ( \delta f^{lm} N^{[b}_{lm} q^{a]}) + E^a
\end{equation}
where the equations of motion terms are: 
\begin{equation}
 E^a \equiv 2 q^b \delta(\g G^a_b )+ q^a\g G_{ij} \delta g^{ij}
\end{equation} 
We have mentioned these results only for the sake of completeness and to show why $P^a_H,J^a_H$ etc. are, in fact, the better mathematical constructs. We shall not pursue $P^a_{\rm quad}$ and $J^a_{\rm quad}$ because of their noncovariant nature and associated difficulties.

\subsection{Noether charge for $\zeta^a$}\label{sec:7.3}

Another possible variation of our theme is to use the vector 
$\zeta^a=\xi^a+N^a$ rather than $\xi^a$ as the time development vector. This is indeed a possible choice (because of the $\zeta^a\nabla_at=1$ condition) and, in fact, has been explored in detail in the past literature. We, however, find that  the thermodynamic interpretation is clearer and simpler in terms of $\xi^a$ rather than $\zeta^a$, except when they coincide. Mathematically, the main reason for this  is the fact that $u_a$ is hypersurface orthogonal. Physically, observers traveling on the integral curves of $\zeta^a$ are boosted with respect to the fundamental observers moving orthogonal to the hypersurface. This boost affects their view of physics in terms of modified acceleration, energy density measurements etc and the extra terms  cloud the simple interpretation in terms of the fundamental observers. This is why I have not bothered to discuss this case in detail. However, given the utility of $\zeta^a$ in certain contexts and the popularity it enjoys in the  literature, I will make some brief comments about the Noether charge associated with this vector.

 Just to see the complications which arise, let us attempt a direct computation of the Noether charge for $\zeta^a$. The, by now familiar, Noether current expression
\begin{equation}
 u_a \nabla_b J^{ab}(\zeta) = 2 R^a_b \zeta^b u_a + g^{ij} u_a \pounds_\zeta N^a_{ij}
\end{equation} 
can be re-written in the form 
\begin{equation}
  g^{ij} u_a \pounds_\zeta N^a_{ij} =D_\alpha ( NJ^{\alpha 0})
   -2  R^a_b \zeta^b u_a    
\label{jzeroal}
\end{equation} 
This requires us to compute $NJ^{\alpha 0}$, the integral of which over the boundary will give the Noether charge. Direct computation in the adopted coordinate system gives a rather complicated result:
\begin{equation}
 NJ^{\alpha 0} = 2 N a^\alpha  - \frac{2}{N} N_\beta ( D^\alpha N^\beta) + \frac{1}{N} h^{\alpha\beta} \partial_0 N_\beta
\end{equation} 
When we work in the gauge 
$g_{0\alpha} = N_\alpha =0$ so that $N^a=0$ and $\zeta^a=\xi^a$, we  get back the previous results. Also note that the Komar energy
measured by observers moving on the integral curves of $\zeta^a$ will involve the factor $R_{ab}\zeta^a\zeta^b$ and not $R_{ab}\zeta^a u^b$. 

It is obvious that a more geometrical approach is required for the interpretation of the results in this context. From the standard Gauss theorem 
\begin{equation}
 \int_\mathcal{R} d^3x \, u_a J^a \sqrt{h} = \int_{\partial\mathcal{R}} d^2x \,\sqrt{\sigma}\, u_a r_b J^{ab}
\end{equation} 
we know that we need to compute the expression $r_a u_b J^{ba}$ to calculate the Noether charge. This is straightforward to do and we find that [see Appendix \ref{appen9.2}, \eq{ib5app}]:
\begin{equation}
 r_a u_b J^{ba} = 2 r_a \zeta^b \nabla_b u^a - \left( r_a \pounds_\zeta u^a - u_a \pounds_\zeta r^a\right)
 = 2 r_a \zeta^b \nabla_b u^a -  r_a \pounds_\zeta u^a
\label{ib5} 
\end{equation} 
where the last equality arises if we choose the foliation to satisfy the natural condition $u_a \pounds_\zeta r^a =0$.
Of these two terms in \eq{ib5}, the first term can indeed be interpreted as arising due to an ``acceleration'' which measures the variation of the velocity $u^a$ along the direction of $\zeta^a$. This, in turn, has two components: 
\begin{equation}
 2 A^i \equiv 2 \zeta^b \nabla_b u^i = 2 N a^i + 2 N^b \nabla_b u^i 
\end{equation} 
The first term is what we have been working with all along and is related to  the standard acceleration. The second term arises from the spatial drift of the observers moving along the integral curves of $\zeta^a$, with respect to the fundamental observers. 

We can now go ahead and re-compute all the expressions which we originally calculated with $\xi^a$. We will find  everywhere that there are extra terms arising from this boost. We will also encounter the following further difficulties in the interpretation:
 
(a) If we consider regions bounded by $N=$ constant surfaces, then the standard interpretation of  the Davies-Unruh temperature requires the matching of  the \textit{magnitude} $a\equiv (a_ia^i)^{1/2}$ of the acceleration with its normal component so that one can write $r_\alpha a^\alpha = a$. (The Davies-Unruh temperature is associated with the magnitude of the acceleration, not with any given component; so we need $r_\alpha a^\alpha$ on the surface
--- which arises from the Gauss theorem, on integration --- to be equal
 to  $a$ for the interpretation to work.)  This, obviously, will not work if we use $A^i$ as the relevant acceleration and we again need to add boost dependent correction terms to the  thermodynamic quantities to make the interpretation work. 

(b) Second, in relating $\pounds p^{ab}$ to $\pounds N^a_{bc}$ to obtain, say, \eq{niskp}, we use the relation \eq{ib6app} of Appendix \ref{appen9.3}, which contains an extra spatial derivative term. This term identically vanishes for $\xi^a$ but not for $\zeta^a$. One needs to include this contribution in  the holographic equipartition and elsewhere if we use $\zeta^a$. 

(c) We saw in Sec.~\ref{sec:7.1} that the  addition of the  $2K$ term at the boundary does not change the Noether potential if we use $\xi^a$, but it picks up an additional contribution if we use $\zeta^a$.
So the results with the  Einstein-Hilbert Lagrangian and the results for the Lagrangian with an additional $2K$ term will be different when we use $\zeta^a$, but they are the same if we use $\xi^a$.  

As I said before, all these are possible to handle, but I find the description  based on $\xi^a$, a lot more natural and elegant.

\section{Discussion}

Since I have already summarized the results in Sec.~\ref{sec:1} (which the reader is invited to glance through again now!) I will confine myself here to some  comments about the overall perspective.
This will be useful since I have attempted to gather together at least two distinct themes in this paper:

\begin{itemize}
 \item Many results in standard  general relativity, including the evolution equation for spacetime, can be given a thermodynamic interpretation (Sections 3--5 and \ref{sec:7}).
In particular, one can interpret the time evolution as being driven by the departure
from holographic equipartition.
\item The standard formalism ignores the most important \textit{observational fact} about gravitational dynamics and hence is bound to be incorrect. The thermodynamic perspective suggests an alternative paradigm which is closer to observational facts about gravity (Sec.~\ref{sec:6}). 
\end{itemize}

The connecting thread between the two themes is the role played by the combination $f^{ab}\pounds_\xi N^c_{ab}$ throughout the paper. This quantity 
(1) reinforces the utility of the variables $(f^{ab}, N^c_{ab})$ (emphasized in \cite{KBP}),
(2) allows us to introduce gravity though a momentum space Lagrangian,
(3) motivates the choice for the gravitational momentum flux, 
(4) relates to the surface term in the action principle, 
(5) connects up various properties of the Noether current and 
(6) assumes the center-stage in the equation for the evolution of spacetime in terms of the departure from holographic equipartition. Obviously, one needs to understand the physical significance of this term  from different angles (and find it a good name!).

We also found that the interpretations are simpler and more elegant when we use $\xi^a=Nu^a$ as our time development vector. This is a new feature since most of the discussion in the  past literature has concentrated on $\zeta^a=\xi^a+N^a$. I hope to do a more detailed comparison in a future work to clarify their inter-relationship,  but I believe $\xi^a$ is the better choice. Using this vector, we could immediately interpret the Noether charge in the thermodynamic language as the surface heat energy (rather than as entropy). We also could clarify further the role played by the equipartition energy $2TS$ \textit{vis-a-vis} the heat content $TS$, amplifying on the work in \cite{a40}. (As an off-shoot, we also know that, not only $M$ but also $M/2$, have a clear physical meaning and there is no factor 2 problem in the Komar integrals!)

The \textit{mathematics} behind many of the results works because of the following facts: (a) There is a well-known identity (see e.g., page 541 of \cite{tpgravity}) for $R_{ab}u^au^b$ given by \eq{ruu} which relates it to $\nabla_i a^i$ and the time evolution terms through $K_{ab}$ or $p^{ab}$. If we take the field equations to be  $2R_{ab}u^au^b=\mathscr{F}_{ab}u^au^b$, and integrate this identity over all space, we can relate the time-evolution terms to the difference between the flux of acceleration and the Komar energy. (b) Further, the same combination $R_{ab}u^au^b$ occurs in the expression for the Noether charge density $u_iJ^i$, allowing us to introduce the boundary variation term of the action (which occurs in the Noether charge) into the dynamics.

But this is an overly simplistic description of the \textit{physics}! When one works out the expressions, there are several other structural features of gravity which combine together to produce these elegant results:

 First is the fact that the Einstein-Hilbert action is a momentum space action, which is why we could introduce  $(f^{ab}, N^c_{ab})$ and express the boundary variation term of the action --- which occurs in the expression for the Noether current --- in terms of the momentum flow. 

Second, the idea almost works, but not quite --- see the comments towards the end of Appendix \ref{appen9.2} --- with $u^a$ and we \textit{need} to use $\xi^a=Nu^a$.  This has to do with $Na$ being finite on real horizons where $N=0$, in contrast to $a$. (It just does not work with $\zeta^a$ as nicely.) 

Third, the flux of acceleration means nothing in classical general relativity and we need to introduce local Rindler observers and the Davies-Unruh temperature to bring in the thermodynamic perspective. This, of course, has been the corner-stone of the emergent gravity program all along. In addition to the \textit{algebra} of rewriting $(a\sqrt{\sigma}/8\pi)$ as $Ts$, there is an important \textit{conceptual} angle which appears in this approach.
In the standard interpretation of gravity, a field equation like $2\mathbf{G}=\mathbf{T}$, is a geometrical statement independent of any observer. But this equation has the same mathematical content as $2R_{ab}u^au^b=\mathscr{F}_{ab}u^au^b$, if we insist that the latter should hold for all observers with four-velocities $u^i$. It is this equation $2R_{ab}u^au^b=\mathscr{F}_{ab}u^au^b$ ---  or, more precisely, its left hand side --- with an extra four-velocity field in it,  that is amenable to the thermodynamic interpretation. This is because, by its very nature, the Davies-Unruh temperature has an observer dependence and one cannot use it to interpret a purely geometrical equation like $2\mathbf{G}=\mathbf{T}$. But when we demand the validity of the thermodynamic interpretation for \textit{all} observers, we \textit{are}  demanding the validity of $2R_{ab}u^au^b= \mathscr{F}_{ab}u^au^b$ for \textit{all} observers, thereby leading to $2\mathbf{G}=\mathbf{T}$. The algebra is the same, but the physics is quite different.

As regards the second theme of the paper, I have already presented detailed arguments as to why we have no hope of solving the \cc\ problem in the conventional approach. Ignoring the \textit{observed} fact that gravity does not respond to shifts in the matter Lagrangian by a constant is as bad as ignoring the principle of equivalence and trying to describe gravity. This fact also tells us that the \cc\ problem exists even at the tree-level of quantum field theory and issues like the  energy of the vacuum, etc. are red herrings. In the alternative approach, we again use a thermodynamically motivated variational principle (with  $f^{ab}\pounds_\xi N^c_{ab}$ playing a crucial role), and demand the validity of $(2R_{ab}-T_{ab})\ell^a\ell^b=0$ for all null vectors.
This is very similar in spirit to the  idea of demanding  the validity of $2R_{ab}u^au^b=\mathscr{F}_{ab}u^au^b$ for all time-like unit vectors to reproduce standard general relativity. As I explained above, the thermodynamic interpretation requires such an approach. The complete story regarding the \cc\ also requires a principle to fix its numerical value, which is described elsewhere \cite{hptp}.

\appendix

\numberwithin{equation}{section}

\section{Appendices: Calculational details}

Several calculational details and background results are collected together in these appendices in order not to distract the flow of ideas in the main paper. Many of them exist in the literature but are compiled together for the sake of completeness. I have given a fair amount of details (and sometimes alternative derivations of the results) in the hope that these will be useful. Some of these results are easier to obtain in the index-free differential form language but I have presented everything in the more familiar index language.

\subsection{An identity relating the Noether currents}\label{appen9.1}

 It is obvious that  $J^{ab}[q]=0$ identically, if $q_a=\nabla\phi$ is a pure gradient. Very often, we will need to find the Noether current for vectors of the form $v_a=f(x)\nabla_a\phi$. With this motivation,  we will first prove a general relation between Noether currents for two vector fields $q^a$ and $v^a \equiv f(x) q^a$. 
Computing the expanded form of 
\begin{equation}
 J^a(v)=\nabla_b J^{ab}(v) = \nabla_b \left[ \nabla^a (f q^b) - \nabla^b (f q^a)\right]
\end{equation} 
one can easily show that 
\begin{eqnarray}
J^a(v)= \nabla_b J^{ab} ( v) &=& \nabla_b [f J^{ab} (q) + q^b f^a - q^a f^b] = f  J^{a}(q)+J^{ab}(q) f_b  + \nabla_b A^{ba}\nonumber\\
&=& 2 R^a_b f q^b + g^{ij} \pounds_v N^a_{ij}  = g^{ij} \pounds_v N^a_{ij} + f ( J^{a}(q) - g^{ij} \pounds_q N^a_{ij})
\end{eqnarray}
where we use the notation $f_a \equiv \nabla_a f $ and $A_{ij} \equiv q_i f_j - q_j f_i$. This allows us to obtain the relation we need, viz.,
\begin{equation}
J^a(fq)-f  J^{a}(q)=
g^{ij} \pounds_{fq} N^a_{ij} - f g^{ij} \pounds_q N^a_{ij} = f_b J^{ab}(q) + \nabla_b A^{ba} 
\end{equation} 
We  take the dot product of this equation with $q_a$ to obtain
\begin{equation}
 q_a J^a (fq) - f q_a J^a (q) = q_a f_b J^{ab} (q) + q_a \nabla_b A^{ba}
 \label{6nov1}
\end{equation} 
Writing the last term as
\begin{eqnarray}
 q_a \nabla_b A^{ba} &=&  \nabla_b (q_a A^{ba}) - \half A^{ba} J_{ba} (q) = \nabla_b (q_a A^{ba}) - \half \, 2(q^b f^a) J_{ba} (q)\nonumber\\
&=& \nabla_b (q_a A^{ba}) - (q^b f^a) J_{ba} (q)
\label{mid1}
\end{eqnarray} 
we see that the last term in \eq{mid1} cancels with first term in the right hand side of \eq{6nov1}. Therefore, using the definition $A_{ij} \equiv q_i f_j - q_j f_i$ to simplify $\nabla_b (q_a A^{ba})$, we get: 
\begin{equation}
 q_a J^a (fq) - f q_a J^a (q) = \nabla_b (q_a A^{ba})= \nabla_b\left( \left[ q^a q^b - q^2 g^{ab} \right] \nabla_a f\right)
\end{equation} 
In other words, we have the simple result 
\begin{equation}
 q_a J^a (fq) - fq_a J^a(q) =
\begin{cases}
 -\nabla_b(q^2 \mathcal{P}^{ab}(q)\nabla_a f) & \text{when $q^2\ne 0$}\\
+\nabla_b(q^b q^a \nabla_a f) & \text{when $q^2=0$}
\end{cases}
\label{genresult1}
\end{equation} 
where $\mathcal{P}^{ab}(q) = g^{ab} - (q^a q^b / q^2)$ is the projection tensor orthogonal to $q^a$. This result provides one way of computing the  Noether current for vector fields.

\subsection{Noether currents for $u^a$ and $\xi^a$ }\label{appen9.2}

In the  paper, we extensively used the Noether current associated with the vector $\xi_a=Nu_a=-N^2 \nabla_a t$. Here we will compute the Noether current for \textit{both} the  vector fields  $u_a=- N \nabla_at$ and $\xi_a=-N^2 \nabla_a t$ because their comparison shows some structural differences which are of interest (and also because they can be computed without extra effort!).  To calculate these currents, it is convenient to use the identity in \eq{genresult1}. 
We start with $u_a = - N \nabla_at$ and  use \eq{genresult1} with $q_a = - u_a/N,\ f= - N J^a [q]=0$ to get
\begin{equation}
 - \frac{u_a}{N} J^a(u) = - \nabla_b \left(- \frac{1}{N^2} h^{ab} (-\nabla_a N)\right)
 = - \nabla_b (a^b/N)
\label{ib7}
\end{equation} 
where  $h^{ab} = g^{ab} + u^au^b$ is the spatial projection tensor and $h^{ab} \nabla_a N = N a^b$ where $a^b$ is the acceleration of $u^a$. 
To verify that this is indeed the acceleration, even in a general, time-dependent situation, let us compute it directly from the definition $a_k=u^l\nabla_lu_k$. We get:
\begin{equation}
a_k=u^l\nabla_l(-N\nabla_kt)= 
(u^l\nabla_lN)\frac{u_k}{N}+Nu^l\nabla_k\left(\frac{u_l}{N}\right)
=u_ku^l\frac{\nabla_lN}{N}+\frac{\nabla_kN}{N}
=h^l_k\frac{\nabla_lN}{N}
\label{adef1}
\end{equation} 
where we have used $\nabla_k\nabla_lt=\nabla_l\nabla_kt$ to arrive at the  second term in the second equality. This shows that $Na_i=h^j_i\nabla_jN$ which is a useful result.
Expanding out $\nabla_b (a^b/N)$ in \eq{ib7}, we get 
\begin{equation}
 \frac{u^a}{N} J^a(u) = \nabla_b (a^b/N) = \frac{1}{N} \nabla_b a^b - \frac{1}{N^2} (N a^2) = \frac{1}{N} (\nabla_i a^i - a^2)
\label{mid2}
\end{equation} 
To proceed further, we note that, if $v^i$ is any spatial vector  that satisfies the condition $u_iv^i=0$ and $D_i$ is the covariant derivative on  the $t=$ constant surface, we have the result that
\begin{equation}
D_iv^i= D_\alpha v^\alpha \equiv ( g^{ij} + u^i u^j) \nabla_i v_j = \nabla_i v^i - v_j a^j
\label{gen4to3}
\end{equation}
(Note that, in the adopted coordinates we must have $v^0=0$, since $u_i=-N\delta_i^0$.) Applying this to $v^i=a^i$, we have
\begin{equation}
D_ia^i= D_\alpha a^\alpha =  \nabla_i a^i - a^2
\label{di43}
\end{equation} 
Using \eq{di43} in \eq{mid2}, we get the Noether charge density to be:
\begin{equation}
 u_aJ^a[u]=D_\alpha a^\alpha
\label{uju}
\end{equation} 
From the standard expression for the  Noether current we also have:
\begin{equation}
 D_ia^i = 2 R_{ab} u^a u^b + g^{ij}u_a \pounds_u N^a_{ij} 
\end{equation} 

Let us next consider the Noether current for $\xi^a$. Using again \eq{genresult1} with, say, $q^a = u^a, \ f= N$, we  find that 
\begin{equation}
 u_a J^a (Nu) - N u_a J^a(u) = \nabla_b (h^{ab} \nabla_a N) = \nabla_b (N a^b)
\end{equation} 
Using \eq{uju} we  get:
\begin{equation}
u_aJ^a(\xi)= N  u_aJ^{a}(u)+\nabla_j(Na^j)
=N(D_\alpha a^\alpha+\nabla_ja^j+a^2)=2N\nabla_ja^j=2D_\alpha (Na^\alpha)
\label{fin1A}
\end{equation} 
which is the  final result we are after. Of course, in this particular case, one can also obtain this result by direct computation without using \eq{genresult1}. The direct proof of this relation will proceed as follows:
\begin{eqnarray}
 u_b J^b(\xi) &=& \nabla_i (u_b (\nabla^b \xi^i - \nabla^i \xi^b)) - \frac{1}{2} (\nabla_i u_b - \nabla_b u_i) J^{bi}\\
&=& \nabla_i\left[ u_b (N \nabla^b u^i + u^i\nabla^b N  - (i\leftrightarrow b))\right]\nonumber\\
 && \hskip50pt - (a_i u_b - a_b u_i) \left[ (\nabla^b N) u^i + N \nabla^b u^i)\right] \label{seceq}\\
&=& \nabla_i \left[N a^i + u^i u^b \nabla_b N + \nabla^i N\right] - Na^2 - a_b \nabla^b N \\
&=& \nabla_i \left[N a^i + (g^{ib} + u^i u^b) \nabla_b N\right] - Na^2 - a_b \nabla^b N\\
&=& \nabla_i ( 2 Na^i) - Na^2- a_b \nabla^b N = 2 N \nabla_i a^i + a^i \nabla_i N - Na^2 
\end{eqnarray}
where we have used $\nabla_{[i}\, u_{b]} = a_{[i}\, u_{b]}$ to get \eq{seceq}.
Note that the last two terms, when combined as follows, will cancel each other:
\begin{eqnarray}
 Na^i \left[ \nabla_i \ln N - a_i\right] &=& h^{ij}\nabla_j N \left[ \nabla_i \ln N - a_i\right]\nonumber\\
&=& (\nabla_j N) \left[ \frac{1}{N} h^{ij} \nabla_i N - h^{ij} a_i\right] =0
\end{eqnarray}
From \eq{di43}, we also have
\begin{equation}
 \frac{1}{2N}D_\alpha(2N a^\alpha)=D_\alpha a^\alpha+a^\alpha D_\alpha(\ln N)
=D_\alpha a^\alpha+a^2=\nabla_ia^i
\end{equation} 
so that $D_\alpha(2N a^\alpha)=2N \nabla_ia^i$.
This gives the result we are after.

Comparing with the Noether charge densities for $u^i$ (in \eq{uju}) and $\xi^a$ (in \eq{fin1A}) we see that $a^\alpha$ has been replaced by $2Na^\alpha$ which turns out to be important for the  thermodynamic interpretation for two reasons. (i) The factor 2 is crucial in giving the correct expression for  the equipartition result through \eq{flux3}. (ii) More importantly, the combination $Na$ leads to a finite result on the horizon --- leading to standard surface gravity --- while the magnitude $a$ diverges on an $N=0$ surface on static spacetimes. This is also seen from the fact that, the correct local temperatures are always defined 
with the Tolman redshift factor $N$.

Finally, we give the steps involved in proving the expression for  the Noether charge associated with $\zeta^a$ which we needed in Sec.~\ref{sec:7.3}.
It is slightly more convenient to expand the right hand side of \eq{ib5} and show that it is equal to the left hand side. The computation is straightforward: 
\begin{eqnarray}
 2 r_a \zeta^b \nabla_b u^a - ( r_a \pounds_\zeta u^a - u_a \pounds_\zeta r^a) &=& 2r_a \zeta^b \nabla_b u^a - r_a ( \zeta^b \nabla_b u^a - u^b \nabla_b \zeta^a) + u_a (\zeta^b \nabla_b r^a - r^b \nabla_b \zeta^a) \nonumber\\
 &=& r_a u_b \nabla^b \zeta^a -  u_a r_b \nabla^b \zeta^a
 = r_a u_b ( \nabla^b \zeta^a - \nabla^a \zeta^b)\nonumber\\ 
 &=& r_a u_b J^{ba}
\label{ib5app}
\end{eqnarray} 
where we have used the result $r_au^a=0$ to obtain the second equality. This is the result quoted in the text.

\subsection{Different expressions for  $u_a g^{ij} \pounds_\xi N^a_{ij}$}\label{appen9.3}

We will  obtain here the two other equivalent expressions for $u_a g^{ij} \pounds_\xi N^a_{ij}$ used in the text.   
Using $D_\alpha(2N a^\alpha)=2N \nabla_ia^i$ in \eq{jforxi} we get
\begin{equation}
 u_a g^{ij} \pounds_\xi N^a_{ij}  
 = 2 N (\nabla_i a^i - R_{ab} u^a u^b)
\label{diva}
\end{equation}
 On the other hand, from the standard identity for $R_{ab} u^a u^b$ (see e.g., page 541 of \cite{tpgravity}) we have:
\begin{equation}
 R_{ab} u^a u^b = u^a ( \nabla_l \nabla_a u^l - \nabla_a \nabla_l u^l ) = \nabla_l a^l - K^a_lK^l_a + \nabla_a ( K u^a) + K^2
\label{ruu}
\end{equation} 
where $K_{ij} = - \nabla_i u_j - u_i a_j $ is the extrinsic curvature. So
\begin{equation}
 \nabla_i a^i - R_{ab} u^a u^b = K_{ij} K^{ij} - K^2 - \nabla_a (Ku^a)
\end{equation} 
Substituting into \eq{diva} we get:
\begin{eqnarray}
 \frac{1}{2N} u_a g^{ij} \pounds_\xi N^a_{ij}&=& \nabla_i a^i - R_{ab} u^a u^b = K_{ij} K^{ij} - K^2 - \nabla_a (Ku^a)\nonumber\\ 
&=& K_{ij} K^{ij}  -u^a \nabla_a K
\label{ksquare}
\end{eqnarray} 
This allows a simple physical interpretation for the combination $(K_{ij} K^{ij} - K^2)$ in terms of the Lie derivative of $N^a_{bc}$. 

This result can also be obtained directly from the definitions of various quantities. 
 To do this, we first note that, from the definition of extrinsic curvature,
$K_{lm} = - \nabla_l u_m - a_m u_l$, 
we have the result
\begin{equation}
  K_{lm}  K^{ml} = \left( \nabla_l u_m + a_m u_l \right) \left( \nabla^m u^l + a^l u^m \right)
=(\nabla_l u_m) \, (\nabla^m u^l)
\end{equation} 
Further,  using the notation $\alpha_i\equiv \nabla_i \ln N$, we can write:
\begin{eqnarray}
 \nabla_i ( N u_j) &=& N \alpha_i u_j + N(-K_{ij} - u_i a_j) = - N K_{ij} + N (\alpha_i u_j - u_i a_j)\nonumber\\ 
&=& - N K_{ij} + N (\alpha_i u_j - h^k_j \alpha_k u_i)\nonumber\\
 &=& - N K_{ij} + N (\alpha_i u_j - \alpha_j u_i) - u_iu_ju^k \nabla_k N
\end{eqnarray} 
Taking the symmetric and anti-symmetric parts of this expression we get
\begin{equation}
 J_{ab} (\xi) = 2 N ( \alpha_a u_b - \alpha_b u_a ) ; \qquad S_{ab}(\xi) = - 2 N (K_{ab} + u_a u_b u^k \alpha_k)
\end{equation} 
The rest of the calculation proceeds directly as follows:
\begin{eqnarray}
 u_b (\nabla^b S - \nabla_i S^{bi}) &=& \nabla_b ( S u^b) - \nabla_i (u_b S^{bi}) - (\nabla_b u^b) S + (\nabla_i u_b ) S^{bi}\nonumber\\
&=& \nabla_b ( u^b ( - 2 NK + 2N u^k \alpha_k)) - \nabla_i(2N u^i u^k \alpha_k)\nonumber\\
&& \qquad \qquad  + K (-2NK +2N u^k \alpha_k) + S^{bi} (- K_{ib}-a_b u_i)\nonumber\\
&=&-2 \nabla_b (KN u^b ) - 2 N (K^2 - Ku^k \alpha_k) + 2 N (K_{ib} + a_b u_i) ( K^{bi} + u^b u^i u^j \alpha_j)\nonumber\\
&=& - 2 \nabla_b (NK u^b ) + 2 N (K_{ib} K^{bi} - K^2)  + 2 N K u^k \alpha_k\nonumber\\
&=& 2 N ( K_{ij} K^{ji} -K^2) - 2KN u^b \alpha_k - 2N \nabla_b (Ku^b) + 2 KN u^b \alpha_k\nonumber\\
&=& 2N\left[ K_{ij}K^{ji} - K^2  - \nabla_b (Ku^b)\right] = 2 N \left[ K_{ij} K^{ji} - u^b \nabla_b K\right]
\end{eqnarray}
which agrees with \eq{ksquare} when we use the fact that the left hand side is $u_a g^{ij} \pounds_\xi N^a_{ij}$.

In the text we also related $u_a g^{ij} \pounds_\xi N^a_{ij}$ to $h_{ab}\, \pounds_\xi p^{ab}$ where $p^{ab} \equiv - \sqrt{h} (K^{ab} - h^{ab}K)$ is the momentum conjugate to $h_{ab}$. This result can be obtained as follows. We start with  the standard result (see page 550 of \cite{tpgravity}) used in the variation of the action functional with the $2K$ term:
\begin{equation}
 - g^{ab} u_m \, \delta N^m_{ab} = 2 \delta K + K^{ab} \delta h_{ab}+D_a(h^a_b\delta u^b)
\label{ib6app}
\end{equation} 
which expresses the variational term that appears in the Einstein-Hilbert action in terms of the variation of  the extrinsic curvature. The troublesome term is the one involving the spatial derivative $D_a(h^a_b\delta u^b)$, which, fortunately, vanishes when the variation is due to a diffeomorphism along $\xi^a$. This is easy to see from:
\begin{equation}
D_a(h^a_b\pounds_\xi u^b)= D_a(h^a_b\pounds_\xi (\frac{\xi^b}{N}))
= D_a(h^a_b[(\frac{1}{N})\pounds_\xi (\xi^b)+
\xi^b\pounds_\xi (\frac{1}{N})])=0
\end{equation}
because $\pounds_\xi (\xi^b)=0$ and $h^a_b\xi^b=0$.
As regards the rest of the terms in \eq{ib6app}, an 
explicit calculation of $h_{ab} \delta p^{ab}$ gives
\begin{eqnarray}
 h_{ab} \delta \left[(h^{ab} K - K^{ab})\sqrt{h}\right] &=& K \sqrt{h}\, h^{ab}\, \delta h_{ab} + h_{ab} \sqrt{h} (- \delta K^{ab} + K \, \delta h^{ab} + h^{ab} \, \delta K)\nonumber\\
&=& 3 \sqrt{h}\, (h_{ab} \delta K^{ab} + K^{ab} \delta h_{ab}) - \sqrt{h} \, h_{ab}\delta K^{ab} \nonumber\\
&=&  2 \sqrt{h}\, h_{ab} \delta K^{ab} + 3 \sqrt{h}\,K^{ab} \delta h_{ab}
= 2 \sqrt{h}\,\delta K + \sqrt{h}\,K^{ab} \delta h_{ab} \nonumber\\
&=&   \sqrt{h}\,\left\{ 2\delta K + \,K^{ab} \delta h_{ab}\right\} 
\end{eqnarray} 
It follows that:
\begin{equation}
 \sqrt{h}\, u_c g^{ik} \pounds_\xi N_{ik}^c = -  \, h_{ab}\, \pounds_\xi p^{ab}
\label{niskpapp}
\end{equation} 
This result does not hold for  the diffeomorphism along the  vector $\zeta^a$ and one will obtain an extra surface contribution in that case.

For the variations arising from  the diffeomorphism along $\xi^a$, it is also possible to prove the result in \eq{niskpapp}  directly (without using \eq{ib6app}) as follows:
\begin{eqnarray}
 h_{ab} \pounds_\xi p^{ab} &=& \pounds_\xi \left( h_{ab} \sqrt{h}\, ( K h^{ab} - K^{ab})\right) - p^{ab} \pounds_\xi h_{ab} \nonumber\\
&=& \pounds_\xi ( \sqrt{h}\, 2 K ) - p^{ab} ( - 2 N K_{ab})\nonumber\\
&=& 2 \sqrt{h}\, N u^a \partial_a K + 2 K \frac{1}{2} \sqrt{h}\, h^{ab} (- 2 N K_{ab}) + 2 N p^{ab} K_{ab}\nonumber\\
&=& 2 \sqrt{h}\, N u^a \partial_a K - 2 N \sqrt{h}\, K^2 + 2 N \sqrt{h}\, (K^2 - K_{ab} K^{ab})\nonumber\\
&=& 2 N \sqrt{h}\, ( u^a \partial_a K - K_{ab} K^{ab})
\end{eqnarray} 
In the first equality, we have used the definition of $p_{ab} = \sqrt{h}(K h_{ab} - K_{ab})$; to get the second equality, we have used $\pounds_\xi h_{ab} = - 2N K_{ab}$. This result, when combined with \eq{ksquare}, leads to the desired expression.
We can therefore write the  Noether current  expression as
\begin{equation}
 \sqrt{h}\, u_a g^{ij} \pounds_\xi N^a_{ij} = \sqrt{h}\, D_\alpha (2 N a^\alpha)  - 2 N \sqrt{h}\, R_{ab}\, u^a u^b = - h_{ab} (\pounds_\xi p^{ab})
\end{equation} 
This was the result used in the text.

\subsection{Variation of the gravitational momentum flux $P^a_H$}\label{appen9.4}

We will derive the expression for $ \delta ( \g P^a_H)$, discussed in the text, from first principles. 
We  start with the definition for $P^a$  based on the Einstein-Hilbert action for an arbitrary but fixed $q^a$:
\begin{equation}
P^a_H\equiv L_H q^a + g^{lm} \pounds_q N^a_{lm}=J^a_H-2G^a_bq^b 
\label{defph} 
\end{equation} 
 We next introduce an arbitrary variation of the metric, keeping $q^a$ fixed. (The $q_a$ will change but we will not need it.). The variation of $\g P^a_H$ gives
\begin{eqnarray}
 \delta ( \g P^a_H) &=& q^a \delta(R\g) + \delta(f^{lm} \pounds_q N^a_{lm})\\
&=& q^a \left( R_{ij} \delta f^{ij} - \partial_c (f^{lm} \delta  N^c_{lm})\right) + \delta(f^{lm} \pounds_q N^a_{lm})
+ \pounds_q ( f^{lm} \delta  N^a_{lm}) - \pounds_q ( f^{lm} \delta  N^a_{lm}) \nonumber
\end{eqnarray} 
where, in the second line, we have explicitly added and subtracted $\pounds_q (f^{lm} \delta N^a_{lm})$ which is a trick to get the symplectic structure. This gives
\begin{equation}
  \delta ( \g P^a_H) = \delta(f^{lm} \pounds_q N^a_{lm}) - \pounds_q ( f^{lm} \delta  N^a_{lm}) +  q^a ( R_{ij} \delta f^{ij}) 
+ \pounds_q ( f^{lm} \delta  N^a_{lm})  -q^a \partial_c (f^{lm} \delta  N^c_{lm}) 
\end{equation} 
The combination $Q^a\equiv f^{lm} \delta N^a_{lm}=\g g^{lm} \delta N^a_{lm}$ which appears here is a tensor density of weight one and hence
\begin{equation}
 \pounds_q ( f^{lm} \delta  N^a_{lm}) - q^a \partial_c (f^{lm} \delta  N^c_{lm}) 
= \pounds_q  Q^a - q^a \partial_c Q^c = \partial_c(q^c Q^a - q^a Q^c) 
= \partial_c( f^{lm}\delta N^{[a}_{lm} q^{c]})
\end{equation} 
Therefore,
\begin{eqnarray}
\delta ( \g P^a_H) - q^a R_{ij} \delta f^{ij} &=& \delta ( f^{lm} \pounds_q N^a_{lm}) - \pounds_q(f^{lm} \delta  N^a_{lm}) + \partial_c( f^{lm}\delta N^{[a}_{lm} q^{c]})\nonumber\\
&=& \delta f^{lm}\pounds_q N^a_{lm} - (\pounds_qf^{lm}) \delta  N^a_{lm} + \partial_c( f^{lm}\delta N^{[a}_{lm} q^{c]})
  \label{25nova5}
\end{eqnarray} 
where we have used the fact that $\delta$ and $\pounds_q$ operations on $N^a_{lm}$ commute when $\delta q^a=0$ (see Appendix \ref{appen9.6}). Defining the symplectic form 
\begin{equation}
 \g \omega^a (\delta, \pounds_q) \equiv \delta f^{lm} \pounds_q N^a_{lm} - (\pounds_q f^{lm}) \delta N^a_{lm}
\label{sympform}
\end{equation}
we can write \eq{25nova5} as:
\begin{equation}
 \delta ( \g P^a_H) - q^a R_{ij} \delta f^{ij}=\g \omega^a +\partial_c( f^{lm}\delta N^{[a}_{lm} q^{c]})
\label{varPapp}
\end{equation}
If we consider variations with the background being on-shell (i.e., $G_{ab}=0$ but $\delta G_{ab} \ne 0$), then the second term on the left hand side  of \eq{varPapp} vanishes and we get
\begin{equation}
 \delta ( \g P^a_H) =\g \omega^a +\partial_c( f^{lm}\delta N^{[a}_{lm} q^{c]})
\label{pahos}
\end{equation}
This is a completely covariant equation because $P_H^a$ is a covariant object. On using \eq{defph}, we also find:
\begin{equation}
\delta ( \g [J^a_H-  2G^a_bq^b]) = \g \omega^a +\partial_c( f^{lm}\delta N^{[a}_{lm} q^{c]})
 + q^aR_{ij}\delta f^{ij}
\end{equation}
A slightly different, but equivalent, expression is:
\begin{equation}
  \partial_b \left\{ \delta (\g J^{ab}_H) -  \g g^{lm}\delta N^{[a}_{lm} q^{b]}\right\} = \g\omega^a + 2 \delta (G^a_b q^b \g ) + q^a \g G_{lb}\delta g^{lb}
\end{equation} 
which is valid off-shell. If we now assume that the original spacetime is on-shell with $G_{ab}=0$ but $\delta G_{ab}\neq0$, then:
\begin{equation}
  \partial_b \left\{ \delta (\g J^{ab}_H) -  \g g^{lm}\delta N^{[a}_{lm} q^{b]}\right\} = \g\omega^a + 2 \delta (G^a_b q^b \g )
 \label{deljab1}
\end{equation}

We next compute the Lie derivative of the gravitational energy density. To do this from first principles, we again start from the definition of $P^a_H$ in \eq{defph} (for $q^a = \xi^a$) and obtain from it the result
\begin{equation}
 \sqrt{h} u_a P^a_H[\xi] = - t_a f^{lm} \pounds_\xi N^a_{lm} - \g L_H
\end{equation} 
where $t_a \equiv - (u_a/N) = \delta_a^0$ in the adopted coordinates.
The variation of this expression, on using the standard result for $\delta (\g  L_H)$, gives
\begin{equation}
 \delta \left( \sqrt{h}\ u_a P^a_H\right) = - t_a \delta \left( f^{lm} \pounds_\xi N^a_{lm}\right) - \g G_{ab} \delta \gu ab + \partial_c \left( f^{lm} \delta N^c_{lm}\right)
\end{equation} 
Using the usual trick for obtaining the symplectic structure and introducing a factor $t_a \xi^a =1$ in one of the terms, we get
\begin{eqnarray}
  \delta \left( \sqrt{h}\ u_a P^a_H\right) &=& - t_a\left[ \delta \left( f^{lm} \pounds_\xi N^a_{lm}\right) - \pounds_\xi  \left( f^{lm} \delta  N^a_{lm}\right)\right] 
 -  t_a \pounds_\xi  \left( f^{lm} \delta  N^a_{lm}\right) \nonumber\\
&&\qquad + (t_a \xi^a)\partial_c \left( f^{lm} \delta N^c_{lm}\right) - \g G_{ab} \delta \gu ab 
\end{eqnarray} 
Defining the symplectic form $\omega^a$ as in \eq{sympform}  and using the fact that, for any tensor density $Q^a$ we have the result 
$\pounds_\xi Q^a - \xi^a\partial_c Q^c = \partial_c ( \xi^{[c} Q^{a]})$, we get 
\begin{equation}
  \delta \left( \sqrt{h}\ u_a P^a_H\right) + R_{ab} \delta f^{ab} = \sqrt{h}\, u_a \omega^a - \partial_c \left[ u_a f^{lm} \delta N^{[c}_{lm} u^{a]} \right]
\end{equation} 
Simplifying the second term on the right hand side, we find that 
$u_a f^{lm} \delta N_{lm}^{[c}u^{a]} = -h^c_a f^{lm} \delta N^a_{lm}$
leading to the result: 
\begin{equation}
  \delta \left( \sqrt{h}\ u_a P^a_H\right) + R_{ab} \delta f^{ab} = \sqrt{h}\, u_a \omega^a + \partial_c \left[ h^c_a f^{lm} \delta N^a_{lm} \right]
\label{varPuapp}
\end{equation}
Further, using $R_{ab} \delta f^{ab}= \g G_{ab} \delta \gu ab = -  \g G_{ab} \pounds_\xi \gu ab$ and the Bianchi identities, we can simplify the second term on the left. These together lead to the result
\begin{equation}
 \pounds_\xi \left( \sqrt{h}\, u_a P^a_H(\xi)\right)   = \partial_c \left[ \left( 2 G^c_b \xi^b + h^c_a \gu lm \pounds_\xi N^a_{lm}\right) \g \right]
\label{varPuapp1}
\end{equation} 
This is the result used earlier.

\subsection{Variational formulation based on the null surfaces}\label{appen9.5}

We  used in the text the following result:
\begin{eqnarray}
 R_{ab} \ell^a\ell^b &=& \ell^j (\nabla_i\nabla_j - \nabla_j\nabla_i)\ell^i = \nabla_i(\kappa  \ell^i) - \nabla_i\ell^j\nabla_j\ell^i  - \nabla_j (\ell^j (\kappa + \theta)) + ( \nabla_i\ell^i)^2\nonumber\\
&=& - \nabla_i(\theta\ell^i)- [\nabla_i \ell^j \nabla_j\ell^i - (\nabla_i\ell^i)^2
] \equiv - \nabla_i (\theta \ell^i) - \mathcal{S}
\label{1-1}
\end{eqnarray}
where we have used the standard result $\nabla_i \ell^i = \theta + \kappa$ 
(with $\kappa$ defined through $\ell^i \nabla_i \ell^j = \kappa \ell^j$)
and  defined  $\mathcal{S}$ by,
\begin{equation}
 \mathcal{S}\equiv [\nabla_i \ell^j \nabla_j\ell^i -(\nabla_i\ell^i)^2 ]
\end{equation} 
 When we integrate  expressions over a null surface
with the measure $d\lambda\ d^2 x \sqrt{\sigma} $, we can ignore terms of the kind $\nabla_i(\phi \ell^i )$ for any scalar $\phi$ since they  produce only boundary contributions. To see this we only need to note that, when $\ell^a$ is affinely parametrized,
\begin{equation}
 \nabla_i (\phi \ell^i) = \frac{d\phi}{d\lambda}+\phi \frac{d}{d\lambda}(\ln \sqrt{\sigma}) = \frac{1}{\sqrt{\sigma}} \frac{d}{d\lambda} (\sqrt{\sigma}\,\phi)
\end{equation}
where we have used $\ell^i\nabla_i=d/d\lambda$ and $\nabla_i \ell^i=d( \ln \sqrt{\sigma})/d\lambda$.
 So
  the integral over the null surface of the first term in \eq{1-1} is given by
\begin{equation}
 \int_{\lambda_1}^{\lambda_2} d\lambda\ d^2x\, \sqrt{\sigma}\, \nabla_i ( \theta \ell^i) = \int d^2x \sqrt{\sigma}\, \theta \bigg|_{\lambda_1}^{\lambda_2}
\label{bt1}
\end{equation} 
which is just a boundary contribution which can be ignored.
  We therefore conclude that 
\begin{equation}
 \int_{\lambda_1}^{\lambda_2} d\lambda\ d^2x\, \sqrt{\sigma}\, \left[-2R_{ab} +  T_{ab}\right] \ell^a\ell^b 
=  \int_{\lambda_1}^{\lambda_2} d\lambda\ d^2x\, \sqrt{\sigma}\, [2\mathcal{S} + T_{ab}\ell^a\ell^b ]
\end{equation} 
when we consider variations of the null vectors $\ell^a$ which vanish at the boundaries ($\lambda = \lambda_1, \lambda_2$). It follows from the arguments given in the text that the field equations can be obtained by varying $\ell^a$ in the expression:
\begin{equation}
 Q\equiv \int_{\lambda_1}^{\lambda_2} \frac{d\lambda\ d^2x}{16\pi}\, \sqrt{\sigma}\, [2\mathcal{S} + 16\pi T_{ab}\ell^a\ell^b ]
\end{equation}
where we have reintroduced $16\pi G$ with $G=1$.

We next compute the Noether current for the null vector $\ell^a$ used in the text.
Let $\ell_a$ be a null congruence defining a null surface which may not be affinely parametrized. 
If we take $\ell_a=A(x)\nabla_a B(x)$, then  $\ell^i\nabla_i\ell_j=\kappa \ell_j$ where
$\kappa=\nabla_iA\nabla^iB=\ell^a\nabla_a\ln A$. We can compute the Noether current for $\ell_a$ using our \eq{genresult1}
and noting that the Noether current for $q_a = \ell_a/A$ is zero. This gives
\begin{equation}
 \frac{\ell_a }{A} J^a (\ell) = \nabla_b \left(\frac{\ell^b }{A}\ \frac{\ell^a}{A} \nabla_a A\right) = \nabla_b \left( \frac{\ell^b}{A}\,\kappa\right) =  \nabla_b( \kappa\ell^b)\frac{1}{A} - \kappa\ell^b\frac{1}{A^2}\nabla_b A = \frac{1}{A}\left\{ \nabla_b( \kappa\ell^b) - \kappa^2\right\} 
\end{equation} 
It follows that
\begin{equation}
 \ell_a J^a (\ell) = \nabla_b( \kappa\ell^b) - \kappa^2=
\ell^b\nabla_b\kappa + \kappa^2 + \kappa \theta - \kappa^2 = \frac{d\kappa}{d\lambda} + \theta \kappa
\label{4to2app}
\end{equation} 
where we have used $\nabla_a \ell^a = \theta + \kappa$ and $\ell^a \nabla_a = d/d\lambda$. 
The quantity $\nabla_b( \kappa\ell^b) - \kappa^2$ is analogous to the right hand side of \eq{di43} in the case of a null vector. But  the projection to a surface `orthogonal' to a null vector  is not well defined for us to introduce a covariant derivative. Instead, we have to work with a co-null vector $k_a$ defined such that $k^2=0$ and $k_a \ell^b =-1$. Then the projection to the 2-surface is provided by the projection vector $q^a_b = \delta^a_b + k^a\ell_b + k_b\ell^a$ and we can  relate $\mathcal{D}_a (\kappa \ell^a) \equiv q^{ab} \nabla_a (\kappa\ell_b)$ to $\nabla_a(\kappa \ell^a)$. In this computation, it is useful to note the identity $\ell_b \nabla_a(\phi \ell^b)=0$ for any scalar $\phi$. This leads to the result
\begin{equation}
 \mathcal{D}_a (\kappa\ell^a) = \nabla_a (\kappa \ell^a) + k^b[\ell_b (\ell^a \nabla_a \kappa)+\kappa\ell^a \nabla_a \ell_b]
 = \nabla_a (\kappa \ell^a) - \left( \frac{d\kappa}{d\lambda} +\kappa^2\right)
\end{equation} 
Therefore, 
\begin{equation}
 \nabla_a (\kappa \ell^a) - \kappa^2 = \mathcal{D}_a (\kappa\ell^a) + \frac{d\kappa}{d\lambda}
\label{null4to2}
\end{equation} 
These results lead to the expressions used in the text.

\subsection{Lie derivatives of expressions involving connections}\label{appen9.6}

We derive a series of results related to Lie derivatives, which will  be useful in several derivations in the paper. 
The Lie derivative $\pounds_q T $ of some tensor (or tensor density) $T$ 
 has the structure of terms involving $q\partial T$ plus a series of $T\partial q$. So, a variation $\delta (\pounds_q T)$ will involve the sum of terms each having either $T\delta q  $ or $q\delta T$. The sum of all terms involving $q \delta T$ will lead to $\pounds_q (\delta T)$ while the sum of terms with $T\delta q $ will be $\pounds_{\delta q} T$. Therefore, 
\begin{equation}
 \delta (\pounds_q T) = \pounds_q (\delta T) + \pounds_{\delta q} T. 
\end{equation} 
 So, for any tensorial object $T$, the operations $\delta$ and $\pounds_q$ commute if $\delta q =0$.  

This result, however, is not true for non-tensorial objects involving connections, etc. Since we need to handle them, we will derive a series of results for a particular set of them.
To begin with, we have the expression for the
Lie derivative of the connection
\begin{equation}
\pounds_v\Gamma^a_{bc}=\nabla_b \nabla_c v^a+R^a_{\phantom{a}cmb}v^m
\label{lvgamma1}
\end{equation} 
which can be obtained from first principles, knowing the transformation properties of the  connection or from its definition in terms of the metric.
As an aside, we point out the following curious consequence. In flat spacetime, we get a non-zero result:
\begin{equation}
\pounds_v[\Gamma^a_{bc}]_{\rm flat} = \nabla_b \nabla_c v^a 
\label{lie0}
\end{equation} 
(In fact, if we are using Cartesian coordinates, this result tells us that  the Lie derivative of zero is non-zero; i.e, $\pounds_v 0 = \partial_b \partial_c v^a $!). This fact is sometimes used to define a background subtraction scheme in order to make expressions involving connections covariant. The essential idea is to work with the difference $\Gamma^a_{bc}-\Gamma^a_{bc}|_{flat}$, which, of course, is a tensor.
 The result in \eq{lvgamma1} can be re-written as:
\begin{equation}
 \pounds_q \Gamma^a_{bc} = \sp{q}{\Gamma^a_{bc}} + \partial _b \partial_c q^a 
\label{lqG}
\end{equation} 
where $\sp{q}{....}$ stands for the Lie derivative computed treating the indexed object $(....)$ as though it was  a tensor or tensor density. From this, we get the results:
\begin{equation}
 \pounds_q N^a_{bc} = \sp{q}{N^a_{bc}} - \partial _b \partial_c q^a  + \half \left( \delta^a_c \partial_b \partial_l q^l + \delta^a_b \partial_c \partial_l q^l\right) 
 \label{lqN}
\end{equation}
\begin{equation}
\pounds_q(f^{bc} N^a_{bc}) =  \sp{q}{(f^{bc} N^a_{bc})}  - f^{bc} \partial_b \partial_c q^a + f^{ab} \partial_b \partial_l q^l
= \sp{q}{(f^{bc} N^a_{bc})} - \g K^a
\end{equation} 
where 
\begin{equation}
 K^a = g^{lm} \partial_l\partial_m q^a  - g^{al} \partial_l\partial_m q^m
 \label{DefK}
\end{equation} 
For $V^a \equiv - g^{bc} N^a_{bc}$, we have:
\begin{equation}
\pounds_q ( \g V^a) = \sp{q}{\g V^a)} + \g K^a 
\label{25nova3}
\end{equation} 
Further, using $L_{\rm sur} \equiv \partial_a (\g\, V^a)$, we obtain the result:
\begin{equation}
 \pounds_q L_{\rm sur} = \pounds_q  \partial_a(\g V^a) = \pounds_q \partial_a (\g V^a)\Big|_{\rm std} + \partial_a (\g K^a)
\end{equation}
We are now in a position to work out the Lie derivative of 
 $\g\,  L_{\rm quad} \equiv \g\, L_H - L_{\rm sur} $ needed in the paper.  We have
\begin{equation}
 \pounds_q(\g L_{\rm quad}) = \pounds_q(\g L_H - L_{\rm sur})\Big|_{\rm std} -  \partial_a (\g K^a)
= \partial_a \left[ \g L_{\rm quad} q^a \right] -  \partial_a (\g K^a)
\label{ib2app}
\end{equation} 

Further, let us note the following two useful facts: 
(i) Because of the specific structure of \eq{lqN} and \eq{lqG}, \textit{the operations $\delta$ and $\pounds$ still commute} on $\Gamma^a_{bc}$ and $N^a_{bc}$  even though they are non-tensorial.
(ii) Using the notation $\mathcal{V}^a\equiv \g\, V^a$ and  \eq{25nova3}, we see that $\g\, K^a$ disappears in the following combination (which occurs frequently):
\begin{eqnarray}
 \pounds_q ( \mathcal{V}^a) - q^a \partial_c (\mathcal{V}^c) - \g K^a &=& 
\sp{q}{\mathcal{V}^a} - q^a \partial_c ( \mathcal{V}^c) \nonumber\\
&=& q^c\partial_c \mathcal{V}^a - \mathcal{V}^c\partial_c q^a + \mathcal{V}^a\partial_c q^c - q^a\partial_c\mathcal{V}^c \nonumber\\ 
&=& \partial_c( q^c \mathcal{V}^a -  q^a \mathcal{V}^c)
\label{keyres1}
\end{eqnarray} 

These results allow us to relate the Noether currents for the two Lagrangians $L_H$ and $L_{\rm quad}$. Using the standard result for the Noether current for the Einstein-Hilbert action $J^a_H$, we can re-write the result in \eq{defjquad} in the form
\begin{equation}
 \g J^a_{\rm quad} =  \g J^a_H- \left[ \pounds_q ( - \g V^a ) + q^a \partial_c ( \g V^c) + \g K^a\right]
\end{equation} 
Using the notation $\mathcal{V}^a\equiv \g\, V^a$, we see that $\g\, K^a$ disappears in the expression in square brackets because of \eq{keyres1}, giving
 \begin{equation}
J^a_{\rm quad} = J^a_H + \nabla_b ( V^{[a} q^{b]} ) = J^a_H + \nabla_b ( - \frac{f^{lm}}{\g} N^{[a}_{lm} q^{b]})
= J^a_H + \nabla_b ( g^{lm} N^{[b}_{lm} q^{a]})
 \label{25novc5app} 
\end{equation}
where we have used $\mathcal{V}^a = \g\, V^a = - f^{lm} N^a_{lm}$. If we write $J^a_H\equiv J^a_{\rm quad}+J^a_{\rm sur}$, thereby \textit{defining} a Noether current associated with $L_{\rm sur}$ then:
\begin{equation}
 J^{a}_{\rm sur} = - \nabla_b (V^{[a} q^{b]}) = \nabla_b (V^{[b} q^{a]})
=\nabla_bJ^{ab}_{\rm sur}; \quad J^{ab}_{\rm sur}\equiv V^{[b} q^{a]}=q^aV^b-q^bV^a
\end{equation}

\subsection{Variation of the gravitational momentum flux $P^a_{\rm quad}$}\label{appen9.7}

Here we give some of the algebraic details related to the discussion in Sec.~\ref{sec:7.2}. We begin by relating $P^a_H$ and $P^a_{\rm quad}$ by
\begin{eqnarray}
 \g P^a_H &=& f^{lm} \pounds_q N^a_{lm} + q^a L_H \g  \nonumber\\
&=& \pounds_q (f^{lm} N^a_{lm}) - N^a_{lm}\pounds_q f^{lm} + q^a \left(\g L_q -\partial_c (f^{lm} N^c_{lm})\right)\nonumber\\ 
&=& - P^a_{\rm quad} +\pounds_q (f^{lm}N^a_{lm}) - q^a \partial_c (f^{lm} N^c_{lm})
\end{eqnarray} 
Using the notation $\mathcal{V}^a=\g V^a = - f^{lm} N^a_{lm}$ and using \eq{keyres1}, we get
\begin{equation}
 \g P^a_H = - P^a_{\rm quad} - \left[ \pounds_q (\mathcal{V}^a) - q^a \partial_c \mathcal{V}^c\right] 
= - P^a_{\rm quad} - \left[ \g K^a + \partial_c ( q^{[c} \mathcal{V}^{a]})\right]
\label{ib3app}
\end{equation} 
This gives, on combining with the on-shell result in \eq{pahos}, the result:
\begin{eqnarray}
 \delta P^a_{\rm quad} &=& - \delta (\g P^a_H) - \delta (\g K^a) + \partial_c (q^{[c} \delta N^{a]}_{lm} f^{lm} + q^{[c} N^{a]}_{lm} \delta f^{lm})\nonumber\\
&=& -\g \omega^a - \delta (\g K^a) + \partial_c ( q^{[c} N^{a]}_{lm} \delta f^{lm})
\end{eqnarray}
The extra term $\delta(\g K^a)$ arises because of the non-tensorial character of $P^a_{\rm quad}$.

We can also obtain corresponding expressions for $\delta ( \g J^a_{\rm quad})$ by relating it to $\delta (\g J^a_H)$ using  \eq{25novc5}. Then we get:
\begin{eqnarray}
 \delta ( \g J^a_{\rm quad}) &=& \delta ( \g J^a_H) + \partial_b ( \delta f^{lm} N^{[b}_{lm} q^{a]} +  f^{lm}\delta N^{[b}_{lm} q^{a]})\nonumber\\
&=& \delta  f^{lm}\pounds_q  N^a_{lm} -  \pounds_qf^{lm} \delta  N^a_{lm} +  \partial_c( f^{lm}\delta N^{[a}_{lm} q^{c]})+ 2 q^b \delta(\g G^a_b ) \nonumber\\
&& \qquad\qquad\qquad + q^a\g G_{ij} \delta g^{ij} + \partial_b ( \delta f^{lm} N^{[b}_{lm} q^{a]} +  f^{lm}\delta N^{[b}_{lm} q^{a]})\nonumber\\
&=& \g \omega^a + \partial_b ( \delta f^{lm} N^{[b}_{lm} q^{a]}) + E^a
\label{ib4app}
\end{eqnarray}
where the equations of motion terms are: 
\begin{equation}
 E^a \equiv 2 q^b \delta(\g G^a_b )+ q^a\g G_{ij} \delta g^{ij}
\end{equation} 

It is possible to obtain the above result by direct computation as well.  This derivation is instructive in clarifying the origin of the $K^a$ term. Let us start with the definition:
\begin{equation}
 \g\, P^a_{\rm quad} =(N^a_{ij} \pounds_q f^{ij} - L_{\rm quad} q^a\g ) 
\end{equation} 
and compute the on-shell variation $ \delta P^a_{\rm quad}$. To do this, we will first obtain a preliminary result for the on-shell variation of $L_{\rm quad}$ in a specific form. We begin with the standard on-shell result
\begin{equation}
 q^d\delta ( L_{\rm quad} \g) = q^d \partial_c ( N^c_{ab}\ \delta f^{ab})
\end{equation} 
and re-write it in the following way:
\begin{eqnarray}
\label{L3}
  q^d \delta (L_{\rm quad} \g) &=&   q^d \partial_c (N^c_{ab}\, \delta f^{ab})
=   \partial_c (q^d N^c_{ab} \, \delta f^{ab}) -(\partial_c q^d) N^c_{ab} \, \delta f^{ab} \\
&=& \left[ \partial_c (q^c N^d_{ab} \delta f^{ab}) -N^c_{ab} (\delta f^{ab}) \partial_c q^d \right]
+  \partial_c \left[ ( q^d N^c_{ab} - q^c N^d_{ab}) \delta f^{ab}\right]\nonumber
\end{eqnarray}
To arrive at the second line, we have performed the usual symplectic trick of adding and subtracting the quantity $\partial_c (q^c N^d_{ab}\delta f^{ab})$. Let $v^d \equiv N^d_{ab} \delta f^{ab}$, so that
\begin{equation}
 q^d \delta (L_{\rm quad} \g)=[\partial_c ( q^c v^d) - v^c \partial_c q^d]+ 
\partial_c \left[  q^{[d} N^{c]}_{ab} \delta f^{ab}\right]
\end{equation}
Then it is easy to see that (with $\mathcal{V}^a = - f^{lm} N^a_{lm}$)
\begin{eqnarray}
 \pounds_q ( v^d)&=&\pounds_q [-\delta \mathcal{V}^a- f^{ab}\delta N^d_{ab} ]
=-\pounds_q [\delta \mathcal{V}^a]-\pounds_q[f^{ab}\delta N^d_{ab} ]
=\sp{q}{v^d}-\delta (\g K^a)\nonumber\\
&=&\partial_c ( q^c v^d) - v^c \partial_c q^d-\delta (\g K^a)\nonumber\\
&=&\delta f^{ab} \pounds_q N^d_{ab} + N^d_{ab} \delta (\pounds_q f^{ab})
\end{eqnarray}
where we have used \eq{25nova3} to obtain the third equality.
Therefore
\begin{equation}
 \partial_c ( q^c v^d) - v^c \partial_c q^d=[\delta f^{ab} \pounds_q N^d_{ab} + N^d_{ab} \delta (\pounds_q f^{ab})]+\delta (\g K^a)
\end{equation}
where we have used the fact that when $\delta q^a =0$ the Lie differentiation and variation commute giving $\pounds_q \delta (f^{ab}) = \delta \pounds_q(f^{ab}) $.  So
\begin{equation}
  \delta ( q^d L_{\rm quad}\g ) =   \left\{ \delta f^{ab} \pounds_q N^d_{ab} + N^d_{ab}\delta  \pounds_q  f^{ab}\right\} -\partial_c  \left[ q^{[c} N^{d]}_{ab} \ \delta f^{ab}\right] + \delta (\g K^d)
\end{equation} 
 Once we have the variation of $L_{\rm quad}$ in place, we can compute the variation $\delta P^a_{\rm quad}$ defined with $L_{\rm quad}$. We get
\begin{eqnarray}
  \delta P^d_{\rm quad} &=&  \left\{ \delta N^d_{ab} \pounds_q f^{ab} + N^d_{ab} \delta( \pounds_q f^{ab}) -\delta f^{ab} \pounds_q N^d_{ab} - N^d_{ab}\delta \pounds_q  f^{ab} \right\} - \delta (\g K^a)+\partial_c \left[ q^{[c} N^{d]}_{ab}  \delta f^{ab}\right]\nonumber\\
&=&  \left\{ \delta N^d_{ab} \pounds_q f^{ab} - \delta f^{ab} \pounds_q N^d_{ab}\right\} + \partial_c\left[ q^{[c} N^{d]}_{ab}  \delta f^{ab}\right]-\delta (\g K^a)
\end{eqnarray}
This again has a symplectic structure in the first term and an explicit surface term in the second.

\section*{Acknowledgements}
I thank Bibhas Majhi, Krishnamohan Parattu and Dawood Kothawala for extensive discussions and comments on the earlier drafts. I thank Sunu Engineer for discussions. My work is partially supported by the J.C.Bose research grant of the Department of Science and Technology, Government of India.

\end{document}